\documentclass[10pt,pdftex,twocolumn]{article}
\usepackage[margin=1in]{geometry}
\pdfoutput=1

\usepackage[normalem]{ulem}

\usepackage{stmaryrd} 
\usepackage{mathptmx}
\usepackage{mathpartir}
\usepackage{amssymb}
\usepackage{multicol}
\usepackage{makeidx}  

\usepackage{subfig}
\usepackage{graphicx}
\usepackage{amsfonts}

\usepackage{listings}
\usepackage{color}
\usepackage{xspace}
\usepackage{hyperref}
\usepackage{url}
\usepackage{xifthen} 
\usepackage{color}
\usepackage{lmodern}
\usepackage{amssymb} 
\usepackage{paralist}
\usepackage{comment}
\usepackage{tabularx}

\usepackage{parcolumns}
\usepackage{fancyvrb}
\usepackage[final]{fixme}
\usepackage{mathtools}
\usepackage{tikz}
\usetikzlibrary{matrix}
\usepackage[ruled]{algorithm}
\usepackage{algpseudocode}
\usepackage{ioa_code}
\usepackage{macros}

\usepackage{cleveref}
\crefname{section}{\S}{\S\S}
\DeclareCaptionType{copyrightbox}

\usepackage{letltxmacro}
\newcommand*{\SavedLstInline}{}
\LetLtxMacro\SavedLstInline\lstinline
\DeclareRobustCommand*{\lstinline}{%
  \ifmmode
    \let\SavedBGroup\bgroup
    \def\bgroup{%
      \let\bgroup\SavedBGroup
      \hbox\bgroup
    }%
  \fi
  \SavedLstInline
}

\definecolor{dkgreen}{rgb}{0,0.6,0}
\definecolor{gray}{rgb}{0.5,0.5,0.5}
\definecolor{mauve}{rgb}{0.58,0,0.82}

\hypersetup{
    colorlinks=true,
    linkcolor=black,
    citecolor=black,
    filecolor=black,
    urlcolor=black,
}

\VerbatimFootnotes

\definecolor{gpcolor}{rgb}{0.6,0.2,0.3}
\newboolean{showcomments}
\setboolean{showcomments}{true}
\ifthenelse{\boolean{showcomments}}
{ \newcommand{\mynote}[3]{
    \fbox{\bfseries\sffamily\scriptsize#1}
    {\small$\blacktriangleright$\textsf{\emph{\color{#3}{#2}}}$\blacktriangleleft$}}
}
{ 
\newcommand{\mynote}[3]{}}

\newcommand{\contextclass}{contextclass}

\newcommand{\event}{event}
\newcommand{\events}{events}

\newcommand{\context}{context}
\newcommand{\contexts}{contexts}

\newcommand{\aeon}{\textsc{AEON}\xspace}
\newcommand{\aeonso}{\textsc{\ensuremath{\mathrm{AEON_{so}}}}\xspace}

\newcommand{\cservice}{{\bf\it e}Manager}

\newcommand{\Eventwave}{EventWave}
\newcommand{\Orleans}{Orleans}

\newcommand{\users}{users}

\newcommand{\csharp}{C\#}
\newcommand{\eventwave}{EventWave}

\newcommand{\guide}[1]{}

\newcommand{\Programs}{\mathcal{P}}
\newcommand{\aprog}{p}
\newcommand{\Ctxts}{\mathcal{C}tx}
\newcommand{\acontext}[1][]{cx_{ #1}}
\newcommand{\acontextn}[1][]{\lstinline{C}_{ #1}}
\newcommand{\Classes}{\mathcal{C}ls}
\newcommand{\aclass}{cls}
\newcommand{\Methods}{\mathcal{M}}
\newcommand{\amethod}{m}

\newcommand{\Fields}{\mathcal{F}}
\newcommand{\afield}{f}

\newcommand{\bsacontext}[1][]{\lstinline{C}}
\newcommand{\bsacontextt}[1][]{\lstinline{C}'}

\newcommand{\Vars}{\mathcal{V}ar}

\newcommand{\comm}{s}
\newcommand{\Statements}{\mathcal{S}}
\newcommand{\Types}{\mathcal{T}}
\newcommand{\atype}{\tau}

\newcommand{\ctxtdec}[5][\acontext]{\lstinline[basicstyle=\ttfamily\footnotesize]{contextclass}\ #1\ \{\ #3\ #4\ \ifthenelse{\isempty{#5}}{}{#5\ }\}}
\newcommand{\classdec}[3][\aclass]{\lstinline[basicstyle=\ttfamily\footnotesize]{class}\ #1\ \{\ #2\ #3\ \}}
\newcommand{\fielddec}[2][\afield]{#2\ #1}
\newcommand{\methoddec}[4][\amethod]{#2\ #1(#3)\ \{\ #4\ \}}

\newcommand{\avar}{x}
\newcommand{\bvar}{y}

\newcommand{\asyncm}{\mathsf{async}}

\newcommand{\afunc}{dc}

\newcommand{\DCall}{\mathcal{D}Call}

\newcommand{\intt}{\lstinline[basicstyle=\ttfamily\footnotesize]{int}}
\newcommand{\float}{\lstinline[basicstyle=\ttfamily\footnotesize]{float}}

\newcommand{\aval}{v}

\newcommand{\aeid}[1][]{\ifthenelse{\isempty{#1}}{e_{id}}{e_{#1}}}

\newcommand{\acontextset}{\lstinline{G}}

\newcommand{\achildren}[1][]{\acontext_{ch}}

\newcommand{\ret}[1][\aval]{\mathsf{ret}\ifthenelse{\isempty{#1}}{}{(#1)}}

\newcommand{\RNum}[1]{\uppercase\expandafter{\romannumeral #1\relax}}

\begin{document}

\setlength{\pdfpageheight}{\paperheight}
\setlength{\pdfpagewidth}{\paperwidth}

\font\myfont=cmr12 at 16pt

\author{Bo Sang\\
	Purdue University\\
	bsang@purdue.edu\\
	\and
	Gustavo	Petri\\
	IRIF -- Universit\'e Paris Diderot\\
	gpetri@liafa.univ-paris-diderot.fr\\
	\and
	Masoud Saeida Ardekani\\
	Purdue University\thanks{Now at Samsung Research
		America.}\\
	msaeidaa@purdue.edu\\
	\and
	Srivatsan Ravi\\
	University of Southern California\\
	srivatsr@usc.edu\\
	\and
	Patrick Eugster\\
	Purdue University, TU Darmstadt\\
	p@cs.purdue.edu
}

\title{Programming Scalable Cloud Services with AEON}
\date{}

\maketitle

\begin{abstract}
Designing low-latency cloud-based applications that are adaptable
to unpredictable workloads and efficiently utilize modern cloud
computing platforms is hard. The \emph{actor} model is a popular
paradigm that can be used to develop distributed applications: actors
encapsulate state and communicate with each other by sending
events. Consistency is guaranteed if each event
only accesses a single actor, thus eliminating potential data races and deadlocks.  
However it is nontrivial to provide consistency for concurrent events spanning
across multiple actors.
This paper addresses this problem by introducing \aeon: a framework that
provides the following properties:
(i) \emph{Programmability}:
programmers only need to reason about sequential semantics when reasoning about
concurrency resulting from multi-actor events; (ii) \emph{Scalability}:
\aeon runtime protocol guarantees \emph{serializable} and \emph{starvation-free}
execution of multi-actor events, while maximizing parallel execution; (iii)
\emph{Elasticity}: \aeon supports fine-grained \emph{elasticity} enabling the
programmer to transparently migrate individual actors without violating
the consistency or entailing significant performance overheads.

Our empirical results show that it is possible to combine the best of all the
above three worlds without compromising on the application performance.



\end{abstract}

%
%
%
\section{Introduction}
\label{sec:introduction}
Providing cloud-based distributed solutions, and adequately leveraging the various
capabilities provided by cloud providers is pivotal to many modern low-latency
cloud-based applications and services.
However, many of these applications (and services) still follow the de facto
Internet architecture consisting of stateless front and middle tiers,
equipped with a stateful storage tier at the back-end. Since most
services must use this storage back-end, the scalability of the system
as a whole is limited by the latency and throughput of the storage.
To overcome this limitation, it is common practice to add a
caching mechanism. While a caching middle tier might be effective in
enhancing scalability, it comes at the cost of relaxing the
concurrency control provided by the storage back-end. Moreover, this solution fails to
exploit the inherent data locality of the application, since
cache requests need to be shipped to other processes, potentially residing
on a different virtual machine. 

An alternative to the above architecture which has the
potential to overcome these problems is to build a stateful middle
tier using modern programming models based on actors. 
Actors encapsulate state and communicate with each other by sending
\emph{events}. 
In the actor model, consistency is guaranteed if each event
only accesses a single actor, thus eliminating potential data races and deadlocks.  
Yet, this level of abstraction provided by many existing solutions
(e.g., Erlang, Akka) is not appropriate for cloud-based programming
since it is nontrivial to provide consistency for events spanning
across multiple actors.
Typically, the developer using these models still needs to deal with
distributed systems and cloud programming issues such as asynchrony,
failures and deadlock, to mention but a few.

Recent industrial and academic efforts have proposed actor-based
frameworks (e.g., Orleans~\cite{BykovGKLPT11,export:210931} and
EventWave~\cite{ChuangSYGKK13}) for building and deploying
cloud-based services.  For example, Microsoft's Orleans is
being used to implement many services, including Skype and the Halo game
services~\cite{orleansusage}.

All of these frameworks attempt to ensure a subset of the following
properties:
\begin{inparaenum}[(i)] 
\item Programmability: the simplicity of the framework is paramount to
  reduce the learning effort, increase developers' productivity, and
  guarantee the platform adoption.  This aspect can be achieved by
  providing to the programmer the illusion of sequential semantics,
  hence ignoring the consistency challenges that may arise when the service runs
  in the cloud.\label{program}
\item Scalability: to effectively cope with unpredictable workloads,
  the framework -- and in particular its runtime system -- must be
  able to function at different scales;\label{scalab}
\item Elasticity: to achieve an economical solution, the framework
  must be able to automatically scale both in and out by adding and
  releasing resources to adapt to the workload at hand. Moreover, such workload
  adaptation should not violate application invariants or completely stall
  the computation.\label{elast}
\end{inparaenum}

This paper introduces \aeon: a distributed framework for performing Atomic Events over an Ownership Network that addresses
the three concerns above as follows:

(\ref{program}) To achieve \emph{programmability}, \aeon{} enables reasoning
about multi-actor events with sequential semantics in mind.
Specifically, \aeon applications are modeled as a \emph{partially-ordered set} of
dynamically interacting \emph{contexts} that, akin to actors, represent
units of data encapsulation. Our protocol ensures
that all the events are executed in an atomic and strongly consistent
manner (\`a la strict-serializability in transactional systems). In
other words, \aeon{} provides to the programmer the illusion of a
server answering to asynchronous requests one at a time in a
sequential manner.

(\ref{scalab}) Partial-ordering of contexts in \aeon induces an \emph{ownership
network} to organize contexts, whereby access to a context is only granted to
the contexts that directly \emph{own} it.
This partial ordering results in a directed acyclic graph (DAG) of contexts that
is the key for \aeon{} to implement an efficient deadlock-free and
\emph{starvation-free} synchronization protocol. This protocol maximizes
parallel execution of client request events, and is therefore highly scalable. 
This is in stark contrast to the synchronization employed in
Orleans~\cite{BykovGKLPT11,export:210931}, which does not provide strict
serializability, or EventWave~\cite{ChuangSYGKK13} which severely limits
scalability by employing a global synchronization bottleneck.

(\ref{elast}) As foundation for \emph{elasticity}, \aeon{}s runtime
system allows for transparently migrating contexts across different servers of the system
without affecting the semantics of the application, and thus dynamically
adjusts the number of utilized virtual machines to the actual workload.
Specifically, contexts can be automatically distributed
across a data center without exposing the actual location of contexts
in the network (i.e., it enforces \emph{location transparency}~\cite{Karmani2009}). 

We have implemented a highly available and fault-tolerant prototype of
\aeon in \verb!C++!. Our empirical results show that it is possible to combine the best of all three worlds: programmability, scalability and elasticity
  without compromising on the application performance.
Concretely we make the following contributions:
\begin{inparaenum}[\bf (1)] \item After detailing challenges in developing
elastic software in existing state-of-the-art paradigms such as
\Eventwave{}~\cite{ChuangSYGKK13} and Microsoft's
  \Orleans{}~\cite{BykovGKLPT11}, we present a novel programming model
  for building elastic cloud applications in a seamless and effortless fashion
  (\cref{sec:pmodel}).
\item The runtime of \aeon{} implements a novel protocol for executing
  events in a strict serializable and highly scalable manner
  (\cref{sec:runtime})
\item \aeon{'s} runtime supports \emph{customizable} automatic elasticity
through the novel notion of an \emph{elasticity manager} (\cref{migration}).
\item We report an extensive evaluation, where we compare \aeon{}
  against \Eventwave{} and \Orleans{} on Amazon EC2 through a game application
  and the standard TPC-C benchmarks for transactional systems
  (\cref{sec:evaluation}). 
\end{inparaenum}
Related work and final remarks are the subjects of \cref{sec:related} and
\cref{sec:conclusion} respectively.

The implementation along with extended details, including the operational
semantics, are available on the project website:
\url{https://aeon.gitlab.io}

\section{Overview} \label{sec:overview}
In this section, we first identify the challenges with \emph{programming}
support for \emph{scalable} cloud services and applications and
summarize the drawbacks of existing solutions to the problem.
We then provide an overview of \aeon{}, and illustrate how it addresses these
challenges.

\subsection{Existing Work and Drawbacks}
There exist some efforts towards frameworks that help implement scalable
elastic applications while reducing programming effort.
\emph{EventWave}~\cite{ChuangSYGKK13} and \emph{Orleans}~\cite{BykovGKLPT11} are two important works in this space.

\vspace{1mm}\noindent\textbf{Orleans.}
Orleans is an open-source framework, developed by Microsoft, based on
the actor model.
It introduces the concept of \emph{grains}.
Akin to actors, grains are single-threaded.
There are two types of grains: stateful and stateless.
Although Orleans was initially described to support
transactions~\cite{BykovGKLPT11}, the current open-source version does
not provide transactional guarantees.
However, for many cloud applications, transaction(al) execution is
required for correctness since the manual implementation of distributed
transactions always requires considerable effort.
Moreover, it's easy to run into \emph{deadlocks} in Orleans with (a
cycle of) synchronous method calls because general grains are
single-threaded and do not allow reentrance.
Finally, re-distribution of grains is supported in Orleans, but the
migration process provides no guarantees that the application
semantics will be unaffected~\cite{actor-dist}.

\vspace{1mm}\noindent\textbf{EventWave.}
EventWave is the nearest programming model to \aeon{} in which applications are modeled as a \emph{tree} of contexts. EventWave guarantees
strict-serializability by totally ordering all requests at the (single) root
context, assigning an unique id to each request and executing events in order of their ids.
Consequently, EventWave provides only \emph{minimal progress}~\cite{tm-book}.
This clearly limits scalability and overall performance,
as adding more servers provides only limited benefits due to the bottleneck at the tree root.
Moreover, EventWave only provides a simple API for the programmer to manually migrate contexts to specific servers by halting all executions during migration.
This severely hampers elasticity and introduces a nontrivial performance degradation.
EventWave also provides limited programmability since
it organizes contexts strictly as a tree and does not support modification of tree edges.
This prevents programmers from implementing classic distributed data structures such as B-trees and list-sets.
\cref{sec:related} covers the drawbacks of other (perhaps less) related programming models for the cloud.
\subsection{\aeon{} Overview}
Consider a massively multiplayer online (MMO) game, where players can circulate through an arena containing
different buildings and rooms,
each containing different objects.
The players can interact with other players and objects in the same room. Such a MMO game
has to process thousands of concurrent requests in an \emph{asynchronous} environment, thus
emphasizing the need for an efficient protocol
to synchronize client requests.
When there are
too many online players and existing physical servers become contended, new servers must be allocated and some players must be migrated to those. Such players will still be interacting with other players and objects, and so the game service must handle the migrations both quickly and correctly.

Atomic Events and Ownership Network (\aeon{}) is a general programming framework designed precisely to
solve these problems. \aeon{} allows the programmer to write
applications assuming a sequential semantics. The \aeon{} runtime system
efficiently utilizes the distributed computing resources and
supports seamless resource migration without sacrificing the application's correctness, thus relieving
the application programmer of dealing with intricate concurrency issues.

\vspace{1mm}\noindent\textbf{Programmability.}
In Figure~\ref{aeon-implementation} we outline a simplified \aeon{} implementation
of our game. \aeon{} takes an Object-Oriented (OO) approach to implement the
server-side logic -- the structure of the program follows a standard OO
programming approach if we substitute the \lstinline{contextclass} keyword by
\lstinline{class}, except for a few keywords that we will explain shortly
in~\cref{sec:pmodel}. Defining object
structures as \lstinline{contextclass}es instead of regular classes means
that their instances will be automatically distributed, and relocated under
workload pressure by the \aeon{} runtime system as needed.
Notice that the programmer does
not need to implement any additional logic for the application to adapt to
workloads.

For instance, suppose a client wants to put 50 gold coins
into \lstinline{treasure} from \lstinline{gold_mine}. 
To this end, she issues a call of the form
\mbox{\lstinline|event player1.get_gold(50)|}. The only difference
between an event call and a normal remote method call is the
\lstinline{event} call decoration, which indicates to the runtime
system that the call must be executed as an event.
This annotation on the call site (as opposed to the method
declaration) permits the reuse of methods, e.g., \lstinline{get_gold},
both as events for client calls and as conventional synchronous
methods in the case of another context calling it.


\begin{figure}[!t]
 \vspace{-\topsep}
\begin{lstlisting}[label=aeon-implementation, %basicstyle=\footnotesize\ttfamily,
	captionpos=b,frame=b
,escapechar=\%,caption=Simplified game example. Fields of \contextclass{es} are not shown. Red keywords represent new \aeon{} constructs.]
contextclass Building {
  void updateTimeOfDay () {  // change time of day in parallel
    for (Room* room in children[Room])
       async room->updateTimeOfDay();
  }
  readonly int countPlayers() {         // read-only method
    for (Room* room in children[Room])
       count =+ room->nr_players();
    return count;
  }
  ...
}

contextclass Room {
  readonly int nr_players()             // read-only method
    { return children[Player].size(); }
  readonly int nr_items( ) 
    { return children[Item].size(); }
  void updateTimeOfDay() { ... }
  ...
}

contextclass Player {
  int playerId; 
  Item* gold_mine; 
  Item* treasure;
  bool get_gold(int amt) { 
    if( gold_mine->get(amt) ) 
      treasure->put(playerId, amt);
    ...
  }
  ...
}

\end{lstlisting}
\vspace{-1em}
\end{figure}



While asynchronous calls and events have been proposed before,
the \aeon{} programming model relieves the programmer from reasoning about race
conditions, or tediously implementing synchronization mechanisms.
\aeon{} guarantees strict serializability.
Therefore, events change the state of multiple contexts (i.e.,
instances of \lstinline{contextclass}es) even residing on different
machines, while maintaining the appearance of executing atomically and
sequentially.
In our example, an event call to \lstinline{updateTimeOfDay} in a
\lstinline{Building} context updates the time in all of the rooms
before executing any subsequent event.
\vspace{1mm}\noindent\textbf{Scalability.}
In the interest of maximizing \emph{scalability}, the programmer would
like to execute requests from different \users{} 
in parallel.
However, it is not always the case that requests from different \users{}
operate on disjoint data. 
In the case where two or more requests operate on the same data, an
efficient arbitration mechanism must be put in place to avoid strict serializability
violations. Importantly, this mechanism should also avoid the
possibility of deadlocks.

\aeon{} employs a flavor of ownership types (akin to
\cite{BoyapatiLR02,GordonN07} proposed for concurrent programming) to
facilitate parallel yet atomic executions of distributed events:
contexts form a directed acyclic graph (precisely, a \emph{join
  semi-lattice} as detailed in~\cref{sec:pmodel}) structure indicative
of their state sharing.
Two events can run in parallel as long as they do not access shared
portions of state.
In \aeon{}, a simple static analysis guarantees that the context graph
derived from the context-accessibility (i.e. ownership hierarchy)
between different contexts is \emph{acyclic}. In the example, we can
see that a \lstinline{Player} can own any number of \lstinline{Items},
but not vice-versa. 

Assuming two contexts of type \lstinline{Player} sharing a common
child of type \lstinline{Item}, to guarantee the atomic execution
of an event targeting one of the \lstinline{Player} contexts, \aeon{}
delays the execution of events targeting the other \lstinline{Player}
until the former event is terminated. 
Otherwise, the shared \lstinline{Item} context could be the source of
data races, invalidating the serializable execution of both events. 
However, if two events are sent to \lstinline{Players} in different
\lstinline{Rooms}, they can be executed in parallel without violating
strict serializability of the system since they have no shared children.
This enables a high degree of parallelization since a majority of
events sent by different clients do not intersect.

\begin{figure*}[!ht]
\scalebox{.8}{
        \begin{tabularx}{\textwidth}{c|c|c|c}
	~~~~~ & EventWave~\cite{ChuangSYGKK13} & Orleans~\cite{BykovGKLPT11} & \aeon \\ \hline
	Data encapsulation & Contexts & Grains & Contexts \\
	Programmability restraint & Context tree & Unordered grains & Context DAG \\
	Event consistency across actors & Strict serializability & No guarantees & Strict serializability \\ 
	Event progress & Minimal(due to sequential bottleneck)  & Possibility of deadlocks & Starvation-freedom~\cite{HS11-progress}\\ 
	Automatic elasticity & No  & Yes ~\cite{actor-dist} & Yes 
   \end{tabularx}}
\caption{Summary of distributed programming models for building cloud-based stateful applications}\label{fig:main}
\end{figure*}

\vspace{1mm}\noindent\textbf{Elasticity.}
To build a scalable distributed application that caters to dynamic workloads, the programmer would have to
implement logic to:
\begin{inparaenum}[(i)]
\item migrate both data and computation between
  servers in case of a change in the workload;
\item resolve which server has which
  pieces of data at any given time (which is non-trivial given that
  data might migrate);
\item guarantee that ongoing requests are not
  disrupted by migrations.
\end{inparaenum}
Writing even simple applications which meet the desired
scalability criteria would require expert programmers in
distributed systems, and even in that case it would remain an error-prone, time-consuming, and expensive endeavor.
To avoid that such concerns related to distribution outweigh the
concerns related to the actual program logic, 
\aeon{} employs efficient migration protocols together with an \emph{elasticity manager} that
enables the programmer to specify how contexts scale in/out.
In our game example, the elasticity manager can easily move
\lstinline{Room} and \lstinline{Player} contexts to different servers of the system when their
current virtual machines become overloaded.
For example, a player that starts a computation-intensive task might
be migrated to a single virtual machine for the duration of the task.
%
%
%
Figure~\ref{fig:main} summarizes the properties provided by \aeon with
respect to Orleans and EventWave.


\section{Programming Model}
\label{sec:pmodel}
\newcommand{\dc}[1]{#1 d}
\newcommand{\Dc}[1]{#1 D}
\newcommand{\aexp}{e}
\newcommand{\Exps}{\mathcal{E}xp}
In this section we describe the principal programming abstractions
offered by \aeon{}. Let us start by presenting a simplified abstract
syntax of \aeon in Figure~\ref{fig:syntax}. Notice first that \aeon{}
provides class declarations, as well as methods and fields like most
mainstream OO programming languages. In addition, the language provides syntax
for the declaration of \emph{contextclasses}.

\newcommand{\aname}[1]{#1 n}

\begin{figure}[t]
  \centering
  \newcommand{\mystyle}[1]{\small \textup{#1}}
  \[
  \scriptsize
  \begin{array}{@{\hspace{0mm}}r@{\hspace{0.5mm}}r@{\hspace{1mm}}c@{\hspace{2mm}}l}
    \multicolumn{2}{l}{{\mystyle{Variables}}\ \avar, \bvar \in \Vars}
    & 
    \multicolumn{2}{l}{{\mystyle{Expressions}}\ \aexp \in \Exps}\\[2pt]
    \multicolumn{2}{l}{{\mystyle{Method Names}}\ \amethod \in \Methods}
    & 
    \multicolumn{2}{l}{{\mystyle{Field Names}}\ \afield \in \Fields}\\[2pt]
    \multicolumn{2}{l}{{\mystyle{Class Names}}\ \aclass \in \Classes}
    & 
    \multicolumn{2}{l}{{\mystyle{Contextclass Names}}\ \acontextn \in \Ctxts}\\[4pt]

    {\mystyle{Program Def.}}\ &\ \aprog \in \Programs &\ ::=\ &
                                                {\color{red} \overrightarrow{\dc{\acontext}}}\ \overrightarrow{\dc{\aclass}}\
                                                                \lstinline[basicstyle=\ttfamily\footnotesize]|main|(\dots) \{\ \comm\ \}\\[2pt]
    {\mystyle{Contextclass Def.}}\ &\ \dc{\acontext} \in \Dc{\Ctxts} & \ ::=\ &
                                                           {\color{red}
                                                                       \ctxtdec[\acontextn]{}{\underline{\overrightarrow{\dc{\afield}}}}{\underline{\overrightarrow{\dc{\amethod}}}}{}}\\[2pt]
    {\mystyle{Class Def.}}\ &\ \dc{\aclass} \in \Dc{\Classes} & \ ::=\
    & \classdec[\aclass]{\overrightarrow{\dc{\afield}}}{\overrightarrow{\dc{\amethod}}}\\[2pt]
    {\mystyle{Type}}\ &\ \atype \in \Types & \ ::=\ &
                                                      \underline{\acontextn}\; |\; \aclass \;
                                                      |\; \intt \; |\; \float \; |\; \atype[]\; |\; \dots \\[2pt]
    {\mystyle{Field Def.}}\ &\ \dc{\afield} \in \Dc{\Fields} & \ ::=\ & \fielddec[\afield]{\atype}\\[2pt]
    {\mystyle{Method Def.}}\ &\ \dc{\amethod} \in \Dc{\Methods} & \ ::=\ &
                                                                       {\color{red}
                                                                           {\lstinline[basicstyle=\ttfamily\footnotesize]|ro|^?}}\
                                                                           \methoddec[\amethod]{\atype}{\overrightarrow{\atype\ \avar}}{\comm}\\[2pt]
    {\mystyle{Decorated Call}}\ & \afunc \in \DCall &\ ::=\ & 
                                                                     {\color{red}{\lstinline[basicstyle=\ttfamily\footnotesize]|event|}}\ x.g(\vec{\avar})\
                                                                     |\ {\color{red} {\lstinline[basicstyle=\ttfamily\footnotesize]|async|}}\ x.g(\vec{\avar})\\[2pt]
    {\mystyle{Statements}}\ &\ \comm \in \Statements & \ ::=\ & {\color{red} \afunc}\ |\ \dots
  \end{array}
  \]
  \vspace{-3mm}
  \caption{Syntax of \aeon{} (excerpt). Underlined types are only
    allowed in the declarations of context fields and methods, not in
    class declarations.}
  \label{fig:syntax}
\end{figure}

\vspace{1mm}\noindent\textbf{Classes and contextclasses.}
An \aeon{} program comprises a series of
contextclass declarations, a series of class declarations, and a \lstinline|main|
function which starts the execution of the \aeon{} program.  
%
A context (an instance of a contextclass) is a \emph{stateful point of service} that receives and
processes requests either
\begin{inparaenum}[(i)]
\item in the form of \emph{events} from clients, or
\item in the form of \emph{remote method calls} from other contexts.
\end{inparaenum}
At a high level, a context can be considered as a container
object or composite object that can be relocated between hosts. 
Contexts encapsulate local state (in the form of fields)
and functionality (in the form of exported methods or events).
In particular, \aeon{} contexts hide internal data representations,
which can only be read or affected through their methods.

Another aspect that distinguishes contextclass declarations from the
standard class declarations is that types appearing in contextclass field
and method declarations can also contain context-type expressions, underlined as
$\underline{\atype}$ in Figure~\ref{fig:syntax}. 
By inspecting the rule for types, we can see that contextclass names
can thus be used as types only in contextclass level code, but not in
normal classes.
Thus, we vastly simplify the management of references (for example for
\emph{garbage collection}, in that passing an object by value does not
implicitly create new references to contexts), and enable a simple
static analysis to check that ownership respects a DAG structure as we
shall describe shortly.
Note that this restriction may be relaxed in future revisions of \aeon{}.


\vspace{1mm}\noindent\textbf{Context ownership network.} 
In a nutshell, \aeon{} contexts are guarded by an ownership mechanism
loosely inspired by the ones proposed in~\cite{Almeida97,BoyapatiLR02}.
%
The concept of ownership allows \aeon{} to establish a partial order among
contexts (when considered transitively), and thus guarantees deadlock freedom
when executing events.
%
%

\begin{figure}[!t]
  \hspace{-3mm}\includegraphics[scale=.25]{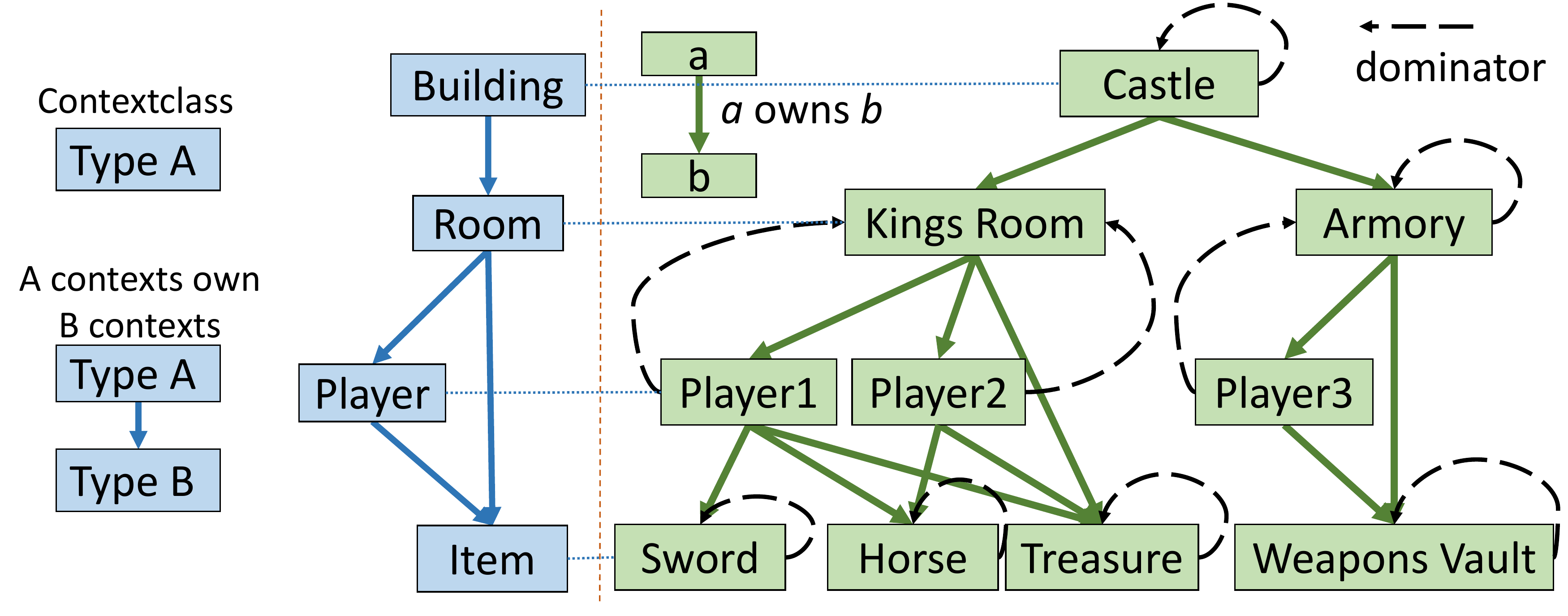}
  \caption{Game static and dynamic context structure.}
  \label{fig:game-context}
\vspace{5pt}
\end{figure}

We say that a context \lstinline{C} is
``directly-owned'' by
another context \lstinline{C}' if any of the fields of \lstinline{C}'
contains a reference to \lstinline{C} (we shall sometimes call the inverse relation
  of directly-owned ``parent-child''). The ownership relation described above takes into
account the transitive closure of the directly-owned relation.
To the right of Figure~\ref{fig:game-context}, we depict a possible runtime
ownership DAG for the application described in Figure~\ref{aeon-implementation}.
Here a \lstinline{Castle} context of type
\lstinline{Building} owns two \lstinline{Room} contexts: the
\lstinline{Kings Room}, and an \lstinline{Armory}. In turn, each of the \lstinline{Rooms} owns the respective \lstinline{Players}
currently in them, and a number of accessible \lstinline{Items}.
\lstinline{Players} can also own \lstinline{Items}. In addition,
some contexts like \lstinline{Treasure} can be owned by multiple
contexts, \lstinline{Player1}, \lstinline{Player2}, and the
\lstinline{Kings Room}. Moreover, several contexts can own the same
context, leading to a form of multi-ownership, which allows the
sharing of state, a prevalent characteristic of object-oriented
programming.

The ownership network enables the safe parallel execution of events
provided that they do not access shared state.  When multiple
concurrent events can potentially access the same state, \aeon{}
serializes the events by exploiting the ownership network.
The DAG structure of the ownership network guarantees that for any two
contexts that might have a common descendant context, there exists an
ancestor context that transitively owns both (we have a
join-semi-lattice).\footnote{Unnamed contexts are automatically added
  in the case of multiple maxima which share common descendants.} In particular, for any
set of contexts that have a common set of descendants, we are
interested in the \emph{least common ancestor} dominating them.
Formally: for context \lstinline{C} in an ownership network
$\acontextset$, assuming that $\mathsf{desc}(\acontextset,\lstinline{C})$
represents the set of its  descendant contexts, 
let $\mathsf{share}(\acontextset,\lstinline{C})$ be the set defined as follows:
\[
\begin{array}{ll}
\footnotesize
\kern-5pt
\mathsf{share}(\acontextset, \bsacontext) = & \kern-7pt \big\{\bsacontextt\ |\ \mathsf{desc}(\acontextset,\bsacontext) \cap
                          \mathsf{children}(\acontextset,\bsacontextt) \neq \emptyset \big\}\ \cup\\
  & \kern-7pt \big\{\bsacontextt\ |\ \mathsf{desc}(\acontextset,\bsacontextt) \cap \mathsf{desc}(\acontextset,\bsacontext) \neq
           \emptyset\ \& \\
  & \kern10pt \ \bsacontextt \notin \mathsf{desc}(\acontextset,\bsacontext)\ \&\ \bsacontext \notin \mathsf{desc}(\acontextset,\bsacontextt)\big\}
\end{array}
\]
Then, we find in $\mathsf{share}(\acontextset, \bsacontext)$ all
contexts which share a descendant context and are otherwise incomparable with
$\bsacontext$ through the directly-owned relation (encoded through
$\mathsf{desc}$), and all the contexts which might be an owner of
$\bsacontext$ and moreover share a common child with \bsacontext. 

In order to calculate the context dominating all contexts that
potentially share something with \bsacontext, denoted
$\mathsf{dom}(\acontextset,\bsacontext)$ and dubbed \bsacontext's ``dominator'',
we can compute the \emph{least upper bound} ($\mathsf{lub}$) of the contexts $\mathsf{share}(\acontextset,\bsacontext) \cup \{\bsacontext\}$ in the lattice $\acontextset$.
\[ 
\footnotesize
\mathsf{dom}(\acontextset,\bsacontext)\ =\ \mathsf{lub}\big(\acontextset,
\mathsf{share}(\acontextset,\bsacontext) \cup \{\bsacontext\} \big) \]

For example, consider Figure~\ref{fig:syntax} which illustrates the ownership
network $\acontextset$ for the game example and indicates dominators for each context:
$\mathsf{dom}(\acontextset,\lstinline{Player1})$ is \lstinline{Kings room} and $\mathsf{dom}(\acontextset, \lstinline{Sword})$ is \lstinline{Sword}.

\vspace{1mm}\noindent\textbf{Methods and events.}  
Events represent asynchronous
client requests to the \aeon{} application, and therefore define its
external API. To simplify the syntactic categories of \aeon{}, and
avoid code duplication, events are simply method \emph{calls}
decorated by the \lstinline{event} keyword targeted at a context. The
same convention applies to asynchronous method calls which are
decorated with the keyword \lstinline|async|.

The execution of events is distributed and can span multiple contexts, but from
the programmers' perspective, the execution of events appears atomic.
The execution of an event conceptually begins at a target
  context: the context providing the method being called.  An event
executing in a certain context \lstinline{C} can issue method calls to any
contexts that \lstinline{C} owns, and in this way can modify the state of any
context transitively reachable in the ownership DAG from \lstinline{C}.

In addition to method calls, events are able to dispatch new events within
themselves.
An event that is dispatched within another event will receive the same treatment
as any other client's event, and will execute after its creator event finishes
its execution.
This is in contrast to synchronous and asynchronous method calls whose execution
is entirely contained within the current event execution.

As shown in Figure~\ref{fig:syntax}, there is an optional
\lstinline|ro| method modifier (\lstinline|ro| is a shorthand for the
more verbose \lstinline|readonly| used throughout). This allows the declaration of methods
that are readonly, which enables the execution of multiple readonly
requests in a single context concurrently. A simple check guarantees that
readonly methods can only use other readonly methods, and that they
cannot modify the state of a context.
In the next section, we will explain in more details how methods and events are
executed.



\vspace{1mm}\noindent\textbf{Type-based enforcement of DAG ownership.}
As stated before, an important invariant to achieve a deadlock-free
strictly serializable semantics for \aeon{} is that the ownership network be
acyclic (at least with respect to contexts that directly export events,
i.e., the entry points for clients to access the application). 


In particular, since the directly-owned relation is related to
referential reachability in the context-graph, we require that the 
graph of contextclasses reachable for a context that exports events be
acyclic. 
An example of a hierarchy is shown in the left hand side
of Figure~\ref{fig:game-context}, where the hierarchy represents
essentially which contextclasses are contained in a certain contextclass.

To enforce this property, we put in place a simple analysis that
collects for each contextclass method declaration, an over-approximation of
the types of contexts that it could access. 
Since our language is in Administrative Normal Form~\cite{FlanaganSDF93}, this
can be done by a single pass over the declarations of contextclasses.
Whenever a contextclass $\acontextn[0]$
declares an event that can use a contextclass $\acontextn[1]$, we
require that the contextclass $\acontextn[0]$ appears always at a
higher level in the ownership network than $\acontextn[1]$ and we
denote this constraint as $\acontextn[1] \leq \acontextn[0]$. The
analysis succeeds if the collected constraints are acyclic except for
the obvious reflexive cases (i.e., $\acontextn[] \leq \acontextn[]$),
and rejects the program otherwise. 
This exception, made for reflexivity of the relation, allows for the
construction of inductive data structures like linked-lists, or trees,
at the slight expense of runtime checks upon modifications of context
ownership structure.
We note that the context ownership structure is modified when the object graph
is explicitly modified. 
%

%

\section{Execution protocol}
\label{sec:runtime} 
\begin{algorithm}[!t]
\caption{ \aeon{} data structures}
\label{alg:ds}

\begin{algorithmic}[1]
  {\footnotesize
    \Part{{\sf Event}}{
      \State  $\ms{eid}$ \quad\Comment{unique event id}
      \State  $\ms{dom}$ \quad\Comment{dominator context}
      \State  $\ms{target}$ \quad\Comment{context the event lands}
    \State  $\ms{accessMode}$ \quad\Comment{indicate readonly or not}
  }\EndPart
  \Statex
  \Part{{\sf Context}}{
    \State  $\ms{cid}$ \quad\Comment{unique id of the context}
    \State  $\ms{toActivateQueue}$ \quad\Comment{queue for incoming events}
    \State  $\ms{toExecuteQueue}$ \quad\Comment{queue for executing events}
    \State  $\ms{activatedSet}$ \quad\Comment{set of events currently
      using the context}
  }\EndPart
}
\end{algorithmic}  
\end{algorithm}

\begin{algorithm}[!ht]
\caption{Event execution at context $C$}
\label{alg:aeon}

\begin{algorithmic}[1]
  {\footnotesize
    
  \Part{{to execute} 
  $\lit{Event}$ $E$
  }{\quad\Comment{accept incoming event}
    \State  $G  \gets \lit{getOwnershipNetwork()}$ \quad\Comment{return context graph}
    \State  $\ms{dom} \gets G.\lit{getDom}(E.target)$ \Comment{get context dominator}
    \State {\bf send} ({\sf ACT}, $E$) to \ms{dom} \Comment{send E to its dom} 
    }\EndPart

\Statex

  \Part{\bf{upon~receive}~({\sf ACT}, {\sf Event} $E$)~\textup{from}~ {\sf Context}~$C'$}{
      \State  $\ms{toActivateQueue}.\lit{enqueue}(E)$
    }\EndPart

\Statex

\Part{{\bf task} $\lit{dispatchEvent}$}{\Comment{dispatch next event}
\While{ $\exists E \in \ms{toActivateQueue}$}
  \State $E \gets \ms{toActivateQueue}.\lit{dequeue}()$
  \State $G  \gets \lit{getOwnershipNetwork()}$
  \If {\big(($\nexists E' \in
  \ms{activatedSet}: E'.\ms{accessMode}=$ {\sf EX}) \&\par
  \hskip\algorithmicindent\; ($E.\ms{accessMode}$ = {\sf RO})\big) } \label{line:ro} \State
  $\ms{activatedSet}$ $\leftarrow$ $\ms{activatedSet} \cup \{E\}$
  \Comment{activate E}
  \Else 
 \State {\bf wait until} $\ms{activatedSet} = \emptyset$
 \State $\ms{activatedSet}$ $\leftarrow$ $\{E\}$ \Comment{activate E}
 \EndIf
 \State {\bf send} ({\sf EXEC}, $E$) {\bf to} $E$.\ms{target} \Comment{send E to execute}
\EndWhile
}\EndPart

\Statex

  \Part{{upon~receive}$~({\sf EXEC}, {\sf Event} ~E)~\textup{from}~{\sf Context}~C'$}{
      \State  $\ms{toExecuteQueue}.\lit{enqueue}(E)$
    }\EndPart
    
\Statex

\Part{{\bf task} $\lit{scheduleNext}$}{ \quad\Comment{scheduling next executing event}
\While{$\exists E \in \ms{toExecuteQueue}$}
  \State  $E \gets \ms{toExecuteQueue}.\lit{dequeue}()$ 
  \If {($E \notin \ms{activatedSet}$)}
  \State $\lit{activatePath}(E)$ \Comment{activate path from target to $C$}
  \EndIf
  \State  $\lit{execute}(E)$ \quad\Comment{execute event after path is activated}
\EndWhile
}\EndPart

\Statex

\Part{{\bf procedure} $\lit{activatePath}$({\sf Event} $E$)}{  \label{alg::aeon::lock}
  \State  $G  \gets \lit{getOwnershipNetwork()}$ \quad\Comment{return context graph}
  \State  $\ms{P} \gets \lit{findPath}(G, E.target, C)$ \Comment{find a path}
  \ForAll{($C' \in p$)} \quad\Comment{activate contexts in the path}
  \State {\bf send} ({\sf ACT}, $E$) {\bf to} $C'$
	\State {\bf wait until} $E$ {\bf is activated at} $C'$
  \EndFor
}\EndPart
}
\end{algorithmic}  
\end{algorithm}

In this section, we describe our novel synchronization protocol employed by
\aeon that arbitrates between two concurrent events to ensure \emph{strict
serializability}: the execution of an application's events built atop \aeon
appears like a \emph{sequential} execution of the application that respects the
\emph{temporal ordering} of events.
In other words, any \aeon{} execution is indistinguishable from a valid
sequential execution of the application' events.
To synchronize among events that execute in contexts that have shared
descendants, \aeon{} employs the dominator context as a sequencer.
Intuitively, when an event is launched in a context \lstinline{C} of an
ownership network \lstinline{G}, the dominator context of \lstinline{C}
(i.e. $\mathsf{Dom}(\lstinline{G},\lstinline{C})$) is conceptually \emph{locked}.
An event locking a context has -- conceptually -- exclusive access to all the
descendants of that context.
Since we lock the dominator context, we have the guarantee that no other event
that shares descendants with \lstinline{C} starts its execution until the
termination of the current event.
These properties are ensured by \aeon{}'s implementation.

\vspace{1mm}\noindent\textbf{Protocol overview.}
Algorithm~\ref{alg:aeon} provides high-level pseudo-code of the \aeon{} synchronization protocol
and Algorithm~\ref{alg:ds} describes the data structures used in Algorithm~\ref{alg:aeon}.
The execution of an \event{} consists of method calls on
the target context, or method calls on contexts that the target
context owns.
To execute, an \event{}  must take the \emph{lock} on the target context.
Each context has a 
set called $\ms{activatedSet}$, which records events that currently lock that
context.\footnote{Multiple readonly events can lock the same context.}
When an event tries to obtain the lock over the dominator of its target context,
it will be placed into the dominator's $\ms{toActivateQueue}$.
When the event tries to lock a \context{} other than the dominator, it
must lock all events in a path from the dominator to itself in a
top-down fashion. 
Finally, when a context method is called, the call is placed in the 
$\ms{toExecuteQueue}$ of the \context{} based on the same ordering determined by
the dominator.




Task $\lit{dispatchEvent}$ dequeues an event from $\ms{toActivateQueue}$, and
waits until the event obtains the lock, that is, it is added to the
$\ms{activatedSet}$.
Note that multiple read-only events can hold the lock to a context at the same
time. 
Once an event takes the lock, it is added to the $\ms{toExecuteQueue}$ for
execution.
Task $\lit{scheduleNext}$ is responsible for dequeuing an
event from $\ms{toExecuteQueue}$, and execute it. 


When a method finishes its execution in a \context{}, control returns
to the caller context, but it is not immediately removed from
the \context{} $\ms{activatedSet}$; the removal happens only when the
\event{} has terminated in all contexts.


The execution model for method calls is by default \emph{synchronous}, similar to Java
RMI. However, in certain situations, for example when notifying every\ children of a
certain context of a change, it is both unnecessary and inefficient to wait for
the completion of a method call before issuing the following one, especially as
these calls may be remote.
The \lstinline!async! method call
decorator in \aeon{} thus indicates that the execution of method call is \emph{asynchronous}.
This is the case of the calls to \lstinline{updateTimeOfDay} for the
\lstinline{Room} contexts in the method of the same name in the declaration of
\lstinline{Building} of Listing~\ref{aeon-implementation}.

Evidently, in the case of multiple asynchronous methods that 
update the state of common children contexts, this behavior can lead to
non-deterministism. 
This is analogous to data races which are considered a
programming error, and have no semantics. In \aeon{} this is also
considered an error, albeit having a well-defined coarse-grained
interleaving semantics (at the level of context accesses). In future
work we will consider ruling out programs prone to
this kind of error at compile-time.

We remark \aeon{} employs a mechanism similar
to read-write locks exploiting the \lstinline{readonly} annotations
(cf. Figure~\ref{fig:syntax}).
Unlike update \events{}, which completely lock the target context,
read-only \events{} conceptually use a \emph{read-lock} (Line~\ref{line:ro}), so multiple read-only
events can execute in parallel in the same context as detailed in Algorithm~\ref{alg:aeon}.

Informally, it is straightforward to see why the \aeon protocol is strictly serializable.
Specifically, let $\mathcal{A}$ be any application built using \aeon and $\pi$ be any execution of $\mathcal{A}$.
There exists a sequential execution $\pi'$ of $\mathcal{A}$ equivalent to $\pi$ such that for any two events $E_1,E_2$ invoked in $\pi$, $E_1 \rightarrow_{\pi} E_2$
implies $E_1 \rightarrow_{\pi'} E_2$, where $E_1 \rightarrow_{\pi} E_2$ denotes the \emph{temporal ordering} between events $E_1$ and $E_2$ in an execution $\pi$.
Indeed, let $\acontextset$ be any ownership network of an application $\mathcal{A}$ and $E_1,E_2$ be any two application events participating in an execution of the \aeon protocol.
Since $\acontextset$ is a semi-lattice (cf. \cref{sec:pmodel}), there is a deterministic monotonic ordering for $E_1$ (and resp. $E_2$) for conceptually locking the contexts accessed 
that begins with $\mathsf{dom}(\acontextset,\lstinline{C})$ (and resp. $\mathsf{dom}(\acontextset,\lstinline{C}')$, where $\lstinline{C}$ (and resp. $\lstinline{C}'$)
is the context on which $E_1$ (and resp. $E_2$) lands initially. Finally, locks on the contexts accessed during an event are \emph{released} in the reverse order on which
they are locked, thus ensuring \emph{starvation-freedom}.

\vspace{1mm}\noindent\textbf{Illustration of event synchronization in \aeon{}.} 
We now illustrate the execution of the \aeon{} synchronization protocol with our game example.
Consider the \emph{single-owner} case where a context \lstinline{C} is its own
    dominator, and an \event{} is enqueued for execution at \lstinline{C}.
    \lstinline{Castle} and \lstinline{Armory} in Figure~\ref{fig:game-context} are
    examples of such single-owner contexts.
In this scenario, to ensure that
no two events modify context \lstinline{C} or its descendants at the same time,
\events{} hold exclusive access of the context hierarchy starting at
\lstinline{C} during their execution, which is guaranteed by enqueuing all
incoming events in the context's execution queue.
Events execute once they reach the head of the queue.

\begin{figure*}[t!]
  \centering
\subfloat{\label{fig:EventExecution-DAG}
  \includegraphics[width=.48\textwidth]{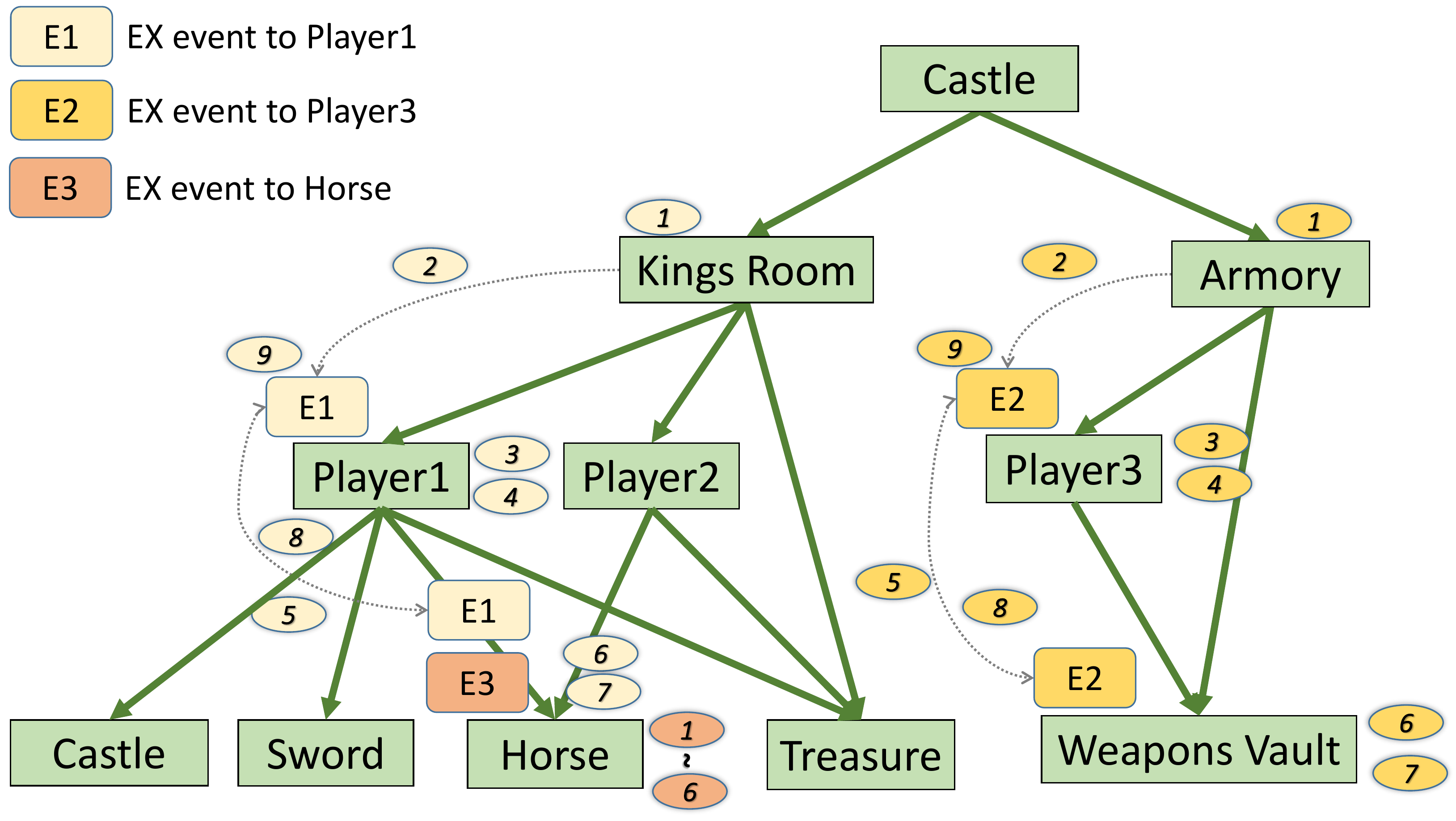}}\hspace{-7pt}
\subfloat{\label{fig:EventExecution-timeline}
  \includegraphics[width=.48\textwidth]{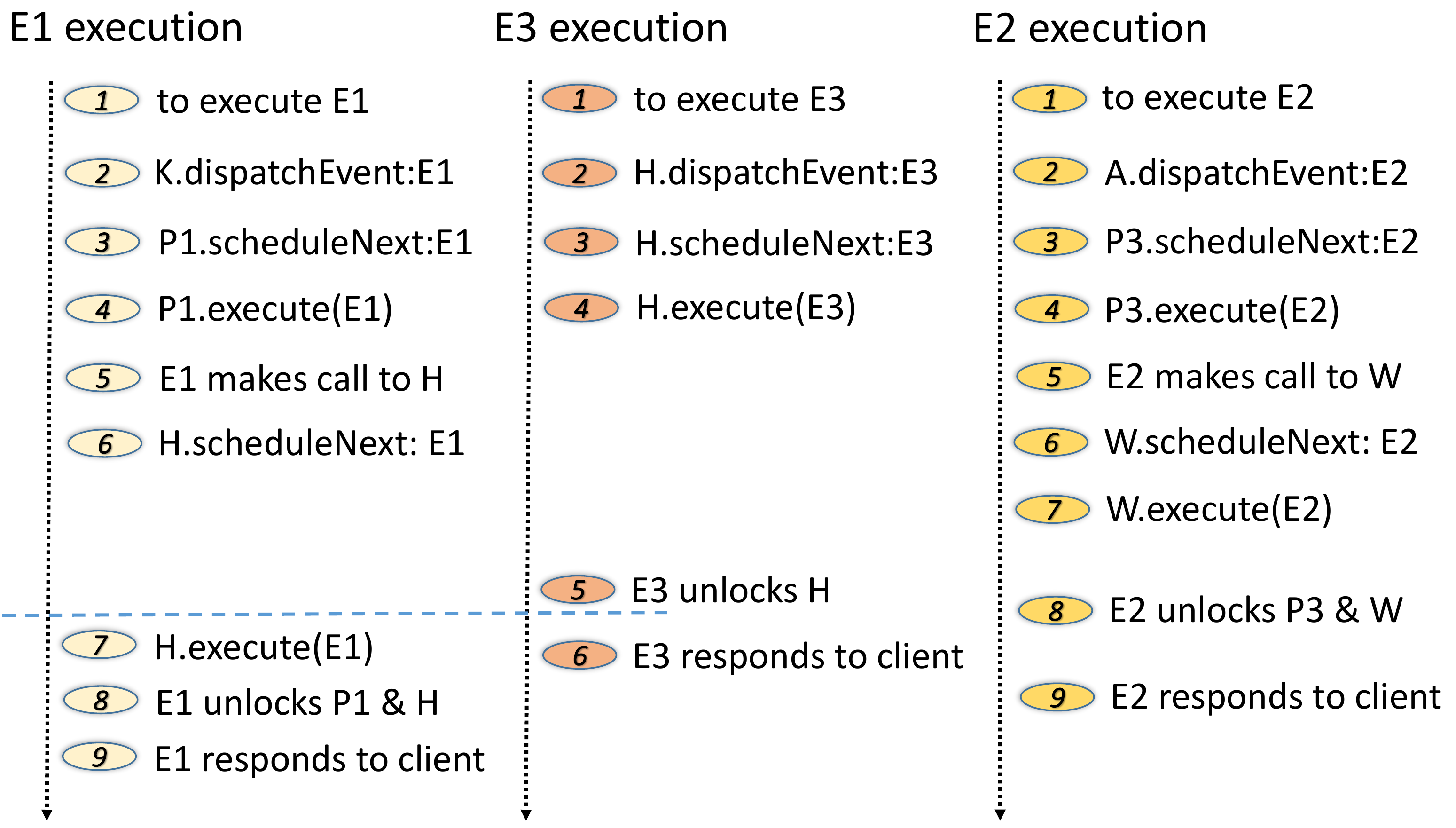}}\hspace{-7pt}
\caption{Event execution}\label{fig:event-execution}
\end{figure*}

However, if there is \emph{context sharing} where the dominator of context \lstinline{C} is
    a different context \lstinline{C}', and the \event{} is forwarded to \lstinline{C}',
    it is sequenced at \lstinline{C}', and starts its execution according to
    the sequence order.
    In Figure~\ref{fig:game-context} all \lstinline{Player}
    contexts are sharing contexts.
The presence of sharing contexts introduces
potential for deadlock.
Consider for example the network in Figure~\ref{fig:game-context}, and
assume that \lstinline{Player1} wants to steal the money from the shared \lstinline{Treasure},
and then run away using the \lstinline{Horse}. At the same time, \lstinline{Player2} wants to
use the \lstinline{Horse} to collect some debts, and then deposit the
money in the \lstinline{Treasure}.
The schema delineated above can lead to a deadlock, where none of the
players is able to execute their events.
To avoid deadlocks, when an \event{} with different target and
dominator contexts is dispatched, \aeon{'s} runtime delivers that
\event{} to the dominator context: the \event{} is serialized at the
dominator context before being sent to its target context for
execution.
Therefore, in Figure~\ref{fig:game-context}, events targeting
\lstinline{Player1} and \lstinline{Player2} need to be
serialized in \lstinline{Kings Room}, whereas events on 
\lstinline{Player3} are serialized in \lstinline{Armory} 
(i.e., their dominator contexts).
Observe that events targeting other contexts can safely execute in
their target contexts.

We observe that for most cases, \contexts{} have different dominators.
For \contexts{} that do not share sub-\contexts{} with others, their dominators
are themselves. Thus events to those \contexts{} will be ordered independently.
Generally, if two \events{} are not ordered by the same dominator, they can execute in parallel.

Figure~\ref{fig:event-execution} shows a timeline of the execution of
three events: $E_1$ targeting context \lstinline|Player1|
(abbreviated P1 in the timeline); $E_2$ targeting \lstinline|Player3|
(P3); and $E_3$ targeting \lstinline|Horse| (H). The numbers in the
timeline correspond to the numbered labels in the ownership graph in
the left of the figure. 

We can firstly observe that since the dominators of $E_1$ and $E_2$
are the \lstinline|Kings Room| and the \lstinline|Armory|
respectively, and these contexts have no common descendants, they can
execute completely independently and in parallel.
The dominator context of event $E_3$ is the \lstinline|Horse|, which
is also its target.
According to the rules outlined above, this event is immediately added
to the $\ms{toActivateQueue}$ of \lstinline|Horse| and subsequently
activated.
Importantly, event $E_1$ also requires to access \lstinline|Horse| in
the timeline.
Therefore, when the execution of $E_1$ reaches the context
\lstinline|Horse|, the $\sf{activatePath}$ procedure will temporarily
stall since in \lstinline|Horse| $E_3$ is currently activated.
Hence, $E_1$ has to wait for the completion of $E_3$ and the
deactivation of $E_3$ in \lstinline|Horse| before resuming its
execution.
In this way, the resulting serialization has the execution of $E_3$
before that of $E_1$, where the latter event sees the effect of the
former one in the context \lstinline|Horse|.


%


%
\newpage
\section{Elasticity} \label{migration}
In this section, we explain \aeon{'s} elasticity manager called \cservice{}.
The \cservice{} provides the following capabilities:
\begin{inparaenum}[(i)]
\item maintaining the global \emph{context mapping} and
  \emph{ownership network}, and
\item managing context creation and migration based on elasticity
  policies.
\end{inparaenum}
In our experiments, \aeon{} is made fault tolerant using the Zookeeper service.
In the remainder of this section, we explain the above two capabilities.

\subsection{Context Mapping}
Since contexts can dynamically migrate across hosts, and in order to deliver
an event to the appropriate context, \aeon{} first needs to find the host
currently holding the corresponding context.
To this end, every client and host caches the most recent context 
mapping that they have queried, and periodically refreshes their context
mappings by querying the \cservice{}.
In practice, and in order to have a highly scalable and available system,
clients and other hosts do not directly query the \cservice{}.
Instead, the \cservice{} stores the latest context mappings along with
the ownership network in a (configurable) cloud storage.
Therefore, to locate a context for the first time (or in case the local cache
has become invalid), a host or a client simply performs a read operation on the
cloud storage system to retrieve the latest mapping.
In the remainder of this paper, and for the sake of simplicity, we assume that
clients and other hosts directly query the \cservice{}.

\subsection{Elasticity Policy}
\aeon{} gives the programmer the ability to define
when and where contexts should be migrated. 
To this end, \aeon{} employs an approach similar to
Tuba~\cite{masoud2014a,masoud2014}. 
Every server periodically sends its resource utilization data (i.e.,
CPU, memory and IO) to the \cservice{}. 
\aeon{} provides a simple API to define when the \cservice{} must
perform a migration.
The following example policies are implemented in \aeon{} by default:
\begin{inparaenum}[(i)]
  \item Resource utilization: in this policy, a programmer defines a lower
  and upper bound of a resource utilization along with an activation threshold.
  Thus, when a resource in a server reaches its upper bound plus a
  threshold the \cservice{} triggers a migration.
  \item Server contention: under this policy, a programmer defines the total
  number of acceptable contexts per server. Hence, once a server reaches its
  maximum, the \cservice{} triggers a migration.
\end{inparaenum} 

Once a migration is triggered, \aeon{} computes a list of possible servers that
can receive the contexts concerned.
The default algorithm tries to move contexts from overloaded hosts to
underloaded ones, but programmers can implement their own algorithms for
choosing hosts and contexts.
In addition, \aeon{} allows programmers to define constraints on any attribute
of the system similar to Tuba~\cite{masoud2014a}. For instance, a
constraint can disallow certain context migrations, or disallow a migration to a new host if
total cost reaches some threshold.


\vspace{1mm}\noindent\textbf{Migration protocol.}
Once a migration is triggered, the \cservice{} will follow the following
\emph{atomic steps} to migrate a context \lstinline{C} from host $s_1$ to a new host
$s_2$.

\begin{compactitem}
\item[\RNum{1}] The \cservice{} sends a prepare message to
  $s_2$, notifying that requests for context \lstinline{C} might start arriving.
  Then, $s_2$ responds by creating a queue for context
  \lstinline{C} and acknowledges the \cservice{}.
  \item[\RNum{2}] 
    Upon receiving the ack, the \cservice{} informs 
    $s_1$ to stop receiving \event{s} targeting \lstinline{C} and it waits for $s_1$
    ack.
  \item[\RNum{3}] 
  Once the \cservice{} receives the ack, and after $\delta$ seconds, it updates
  its context mapping by assigning \lstinline{C} to $s_2$.
Thus, from this point on,
the \cservice{} returns $s_2$ as the location of context \lstinline{C}.
  It then sends a special \event{} called $\lit{migrate}(\lstinline{C},s_2)$ to
  $s_1$ indicating that \lstinline{C} has to be migrated to $s_2$.
  \item[\RNum{4}] 
  Upon receiving $\lit{migrate}(\lstinline{C},s_2)$, $s_1$ enqueues
  an event $\lit{migrate}_c$ in \lstinline{C}'s execution queue. This event serves as a
  notification for context \lstinline{C} that it must migrate.
  When $\lit{migrate}_c$ reaches the head of \lstinline{C}'s queue, $s_1$ spawns a
  thread to move \lstinline{C} to $s_2$.
  \item[\RNum{5}]
   Upon completion of the migration, $s_2$ notifies
  the \cservice{} that the migration is finished, and starts executing
  the enqueued events for context \lstinline{C}.
\end{compactitem}
%
\vspace{1mm}\noindent\textbf{Correctness under context migration.}
Observe that context \lstinline{C}, at the end of step \RNum{2} when $s_1$ stops
accepting \event{s} for \lstinline{C} does not take any steps until step
\RNum{3} when the \cservice{} updates the context map. During this period, $s_1$
does not accept \event{s} targeting context \lstinline{C}, and \cservice{} does not return $s_2$ as the new host for \lstinline{C}.

Once the migration \event{} enters \lstinline{C} at $s_1$ for execution, it will
be the only event that is being executed at \lstinline{C}.
Following the complete execution, both $s_1$ and $s_2$ will have up-to-date
context mappings.
If $s_1$ later receives an event for \lstinline{C} from a host with stale
context map, $s_1$ will forward those \events{} to $s_2$ directly and
notify source host to update its context map. 
In \cref{sec:evaluation}, we will evaluate the performance implication of
halting the execution of events on a migrating context.
%
\subsection{Fault tolerance}
\label{sec:ft}
Similar to Orleans~\cite{BykovGKLPT11}, AEON provides users with a special
\textit{snapshot} API that allows programmers to take consistent snapshots of a
given context along with all its children.
To this end, upon receiving a snapshot request for a context, the
runtime of \aeon{} dispatches a particular event called
$\lit{snapshot}$ to that context.
Consequently, this event takes consistent snapshots of that context and its
children by getting contexts states, and writing them into a (configurable)
cloud storage system like Amazon S3.
To improve the performance, a programmer is able to override a method returning
the state of a context. In case the overridden method returns null for a
context, the runtime system will ignore that context during the checkpointing
phase. 

As we mentioned earlier, in practice the \cservice{} is implemented as a
stateless service that is responsible for updating context mapping and the DAG structure
that are stored in a cloud storage system.
The \cservice{} also leverages the cloud storage system for persisting the steps
of ongoing migrations.
Therefore, if during the course of a migration, the \cservice{}
crashes, a newly elected \cservice{} can read the state of an going migration,
and tries to finish it.
Details on how individual server and the \cservice{} failures are treated
without violating the consistency can be found on the \aeon{} webpage.

\section{Evaluation} 
\label{sec:evaluation} 
We implemented \aeon on top of Mace~\cite{mace}, a \verb!C++! language extension
that provides a unified framework for network communication and event handling.
The implementation of \aeon consists in roughly 10,000 lines of core code and
110 new classes on top of Mace.
In the remainder of this section, we first compare scalability and performance
of \aeon{}  with the two most closely related frameworks:
\eventwave{}~\cite{ChuangSYGKK13} and Orleans~\cite{BykovGKLPT11}.
We then study \aeon{'s} elasticity capabilities, and conclude the section by
evaluating \aeon{'s} migration protocol and its effect on the overall
throughput of the system.

\subsection{Scalability and performance} \label{sec:scalabilityperformance}
In order to compare scalability and performance of \aeon{} with \eventwave{} and
Orleans, we focus on the following two conventional metrics: (i) scaling out:
how a system scales out as we increase the number of servers; and (ii)
performance: how throughput changes with respect to latency as we increase the
number of clients.
To evaluate the above metrics, we implemented the TPC-C benchmark~\cite{tpcc}
and game application in all three systems.

To better study the effect of multiple ownership, the above two applications
were implemented with and without multiple ownership. 
Throughout this section, we refer to the implementation with multiple ownership
as \aeon{}, and refer to the one without multiple ownership as
\aeonso{} (for \textbf{S}ingle \textbf{O}wnership). 
Therefore, the programming effort for implementing the above applications is
identical for \aeonso{} and \eventwave{}. 
 
We run \aeon{}, \aeon{}$_{so}$ and \eventwave{} on m3.large Linux VMs on EC2.
For Orleans and Orleans*, we used m3.large Windows 2012 VMs on EC2.

\begin{figure*}[t]
  \centering 
  \subfloat[Game Scaling out]{\label{fig:Game:Game-scaleout}
    \includegraphics[width=.95\columnwidth]{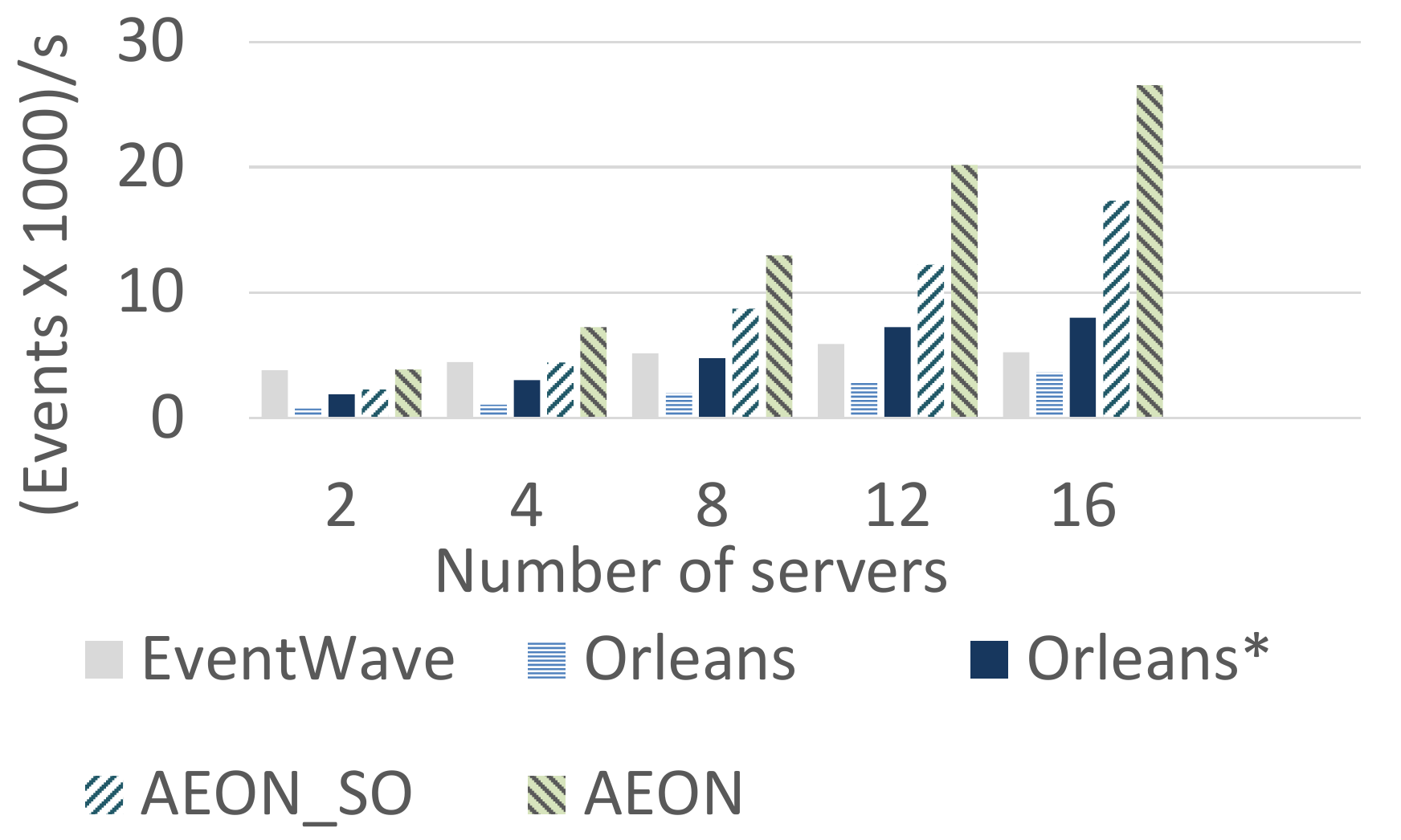}}\hspace{10pt}
  \subfloat[Game Performance]{\label{fig:Game:Game-performance}
    \includegraphics[width=.95\columnwidth]{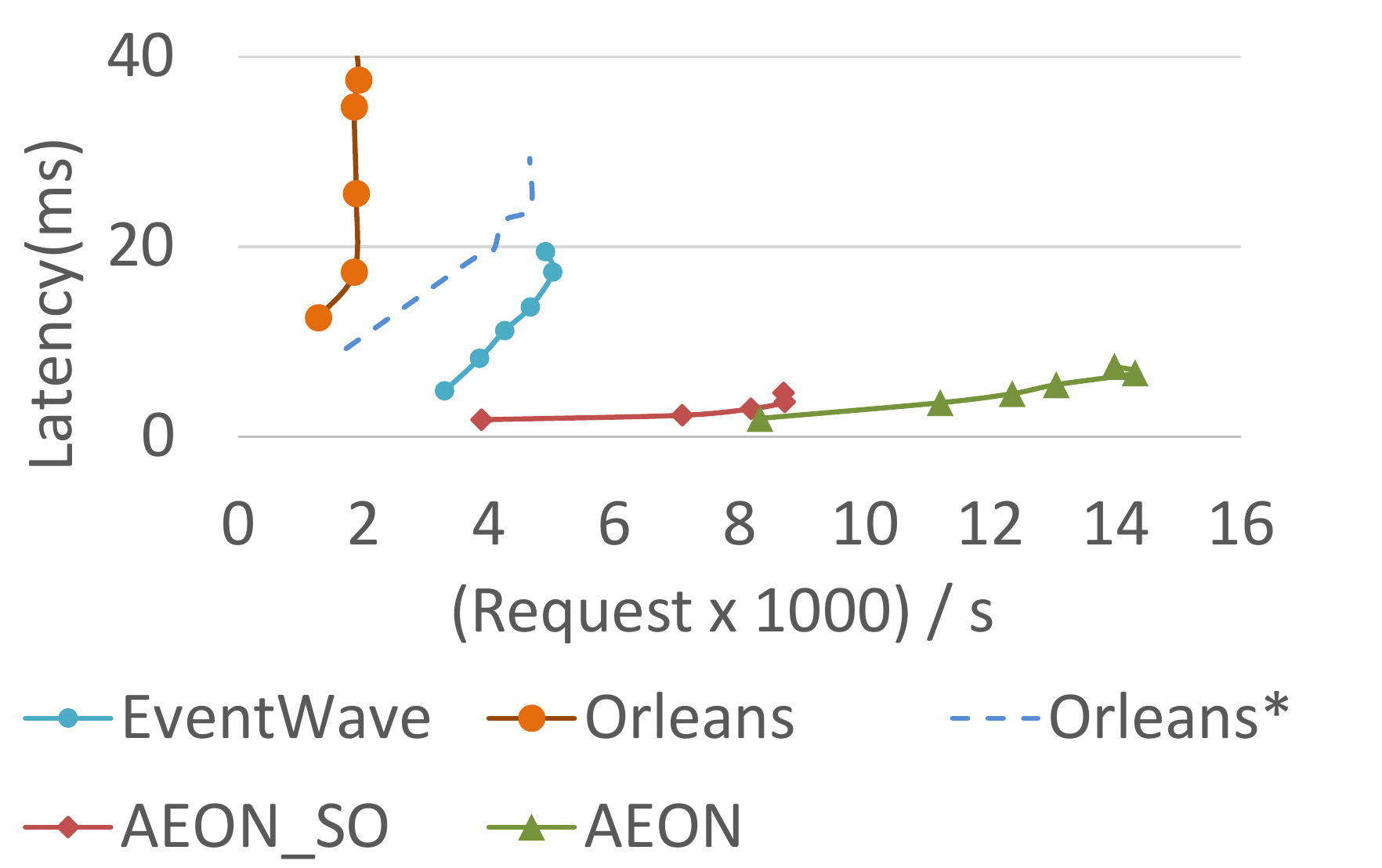}}
  \caption{Game application scalability and performance}\label{fig:Game}
\end{figure*}

\subsubsection{Game application} \label{sec:scalabilityperformance::game} 
Both \eventwave{} and Orleans were previously evaluated using a game application
similar to the example of \cref{sec:overview}.
Therefore, we picked the very same game application described in
\eventwave{}~\cite{ChuangSYGKK13}.
Since \eventwave{} does not support multiple ownership, the implementation does
not allow \lstinline{Players} to access \lstinline{Items} directly. They could
only access \lstinline{Items} via \lstinline{Room}.
Since Orleans doesn't support transactional execution across multiple grains, we
implemented two variants of game application in Orleans:
\begin{inparaenum}[(i)]
  \item A version that ensures strict serializability. This version ensures
  \lstinline{Players} access the shared \lstinline{Items} atomically by
  means of locks. 
  The \lstinline{Players} simply lock the whole \lstinline{Room} when
  they access their \lstinline{Items}.
  This version is called \emph{Orleans} in this section.
\item Since one may argue that the above implementation is not the
  best possible algorithm for implementing the game in Orleans, we also
  implemented a non-strict serializability variant of the game called
  \emph{Orleans*}, in which \lstinline{Players} just access their shared
  \lstinline{Items} directly, and without synchronizing with other
  \lstinline{Players} that have the same \lstinline{Items}.
  This may result in incorrect executions potentially breaking
  application invariants.
  We note that this implementation is only used as a best-case
  scenario for the performance of Orleans, and it should otherwise be
  considered erroneous.
\end{inparaenum}

\vspace{1mm}\noindent\textbf{Scale out.}~ Figure~\ref{fig:Game:Game-scaleout}
shows scalability of different systems for the game application.
In this experiment, we make each server hold one \lstinline{Room} with fixed
number of \lstinline{Items}.
So if there are more \lstinline{Players} in one \lstinline{Room},
\lstinline{Items} will be shared by more \lstinline{Players}.

As shown in Figure~\ref{fig:Game:Game-scaleout}, \eventwave{} reaches maximum
throughput with 12 servers since it needs to order all events in the root node.
Observe that \aeonso{} (resp. \aeon{}) outperforms \eventwave{} by 3x (resp. 5x)
when the number of servers reaches 16.
Since both \aeonso{} and \eventwave{} ensures strict serializability, and have
identical tree structures, the 3x performance gain is not related to multiple
ownership.
Instead, the fact that in \aeon{} events are not ordered at the root context along
with async method calls lead to the observed substantial performance boost.

Interestingly, both \aeon{} and \aeonso{} outperform Orleans* as well.
This is because:
1) \aeon{} is implemented in \verb!C++! and Orleans uses \csharp{}. Hence, we expect
\aeon{'s} implementation to have less overhead.
2) with the help of the ownership DAG, the runtime of \aeon{} can optimize contexts
placement, which will put \lstinline{Rooms}, \lstinline{Players} and
\lstinline{Items} in the \lstinline{Room} on the same server.
Orleans does not have similar rules, which may result in more message passing
among servers.
 3) due to the single-threaded nature of Orleans' grains,
shared \lstinline{Items} have to process requests from \lstinline{Players}
one by one. Though requests could be executed in parallel in
\lstinline{Players}, throughput is limited by the fixed number of
\lstinline{Items} within one Room.
\begin{figure*}[!t]
  \centering
  \subfloat[TPC-C Scaling out]{\label{fig:TPCC-scaleout}
    \includegraphics[width=.95\columnwidth]{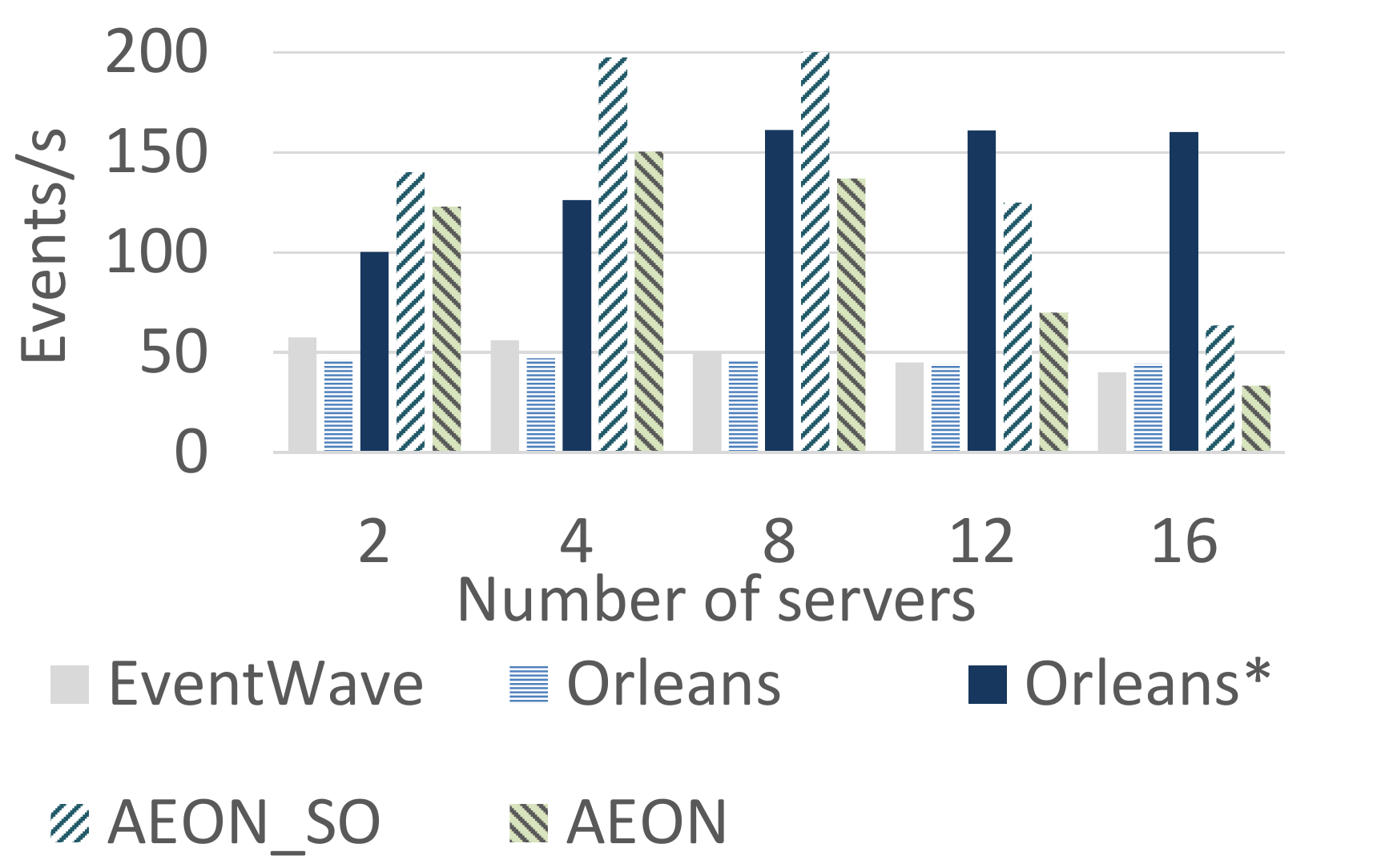}}\hspace{10pt}
  \subfloat[TPC-C Performance]{\label{fig:TPCC-performance}
    \includegraphics[width=.95\columnwidth]{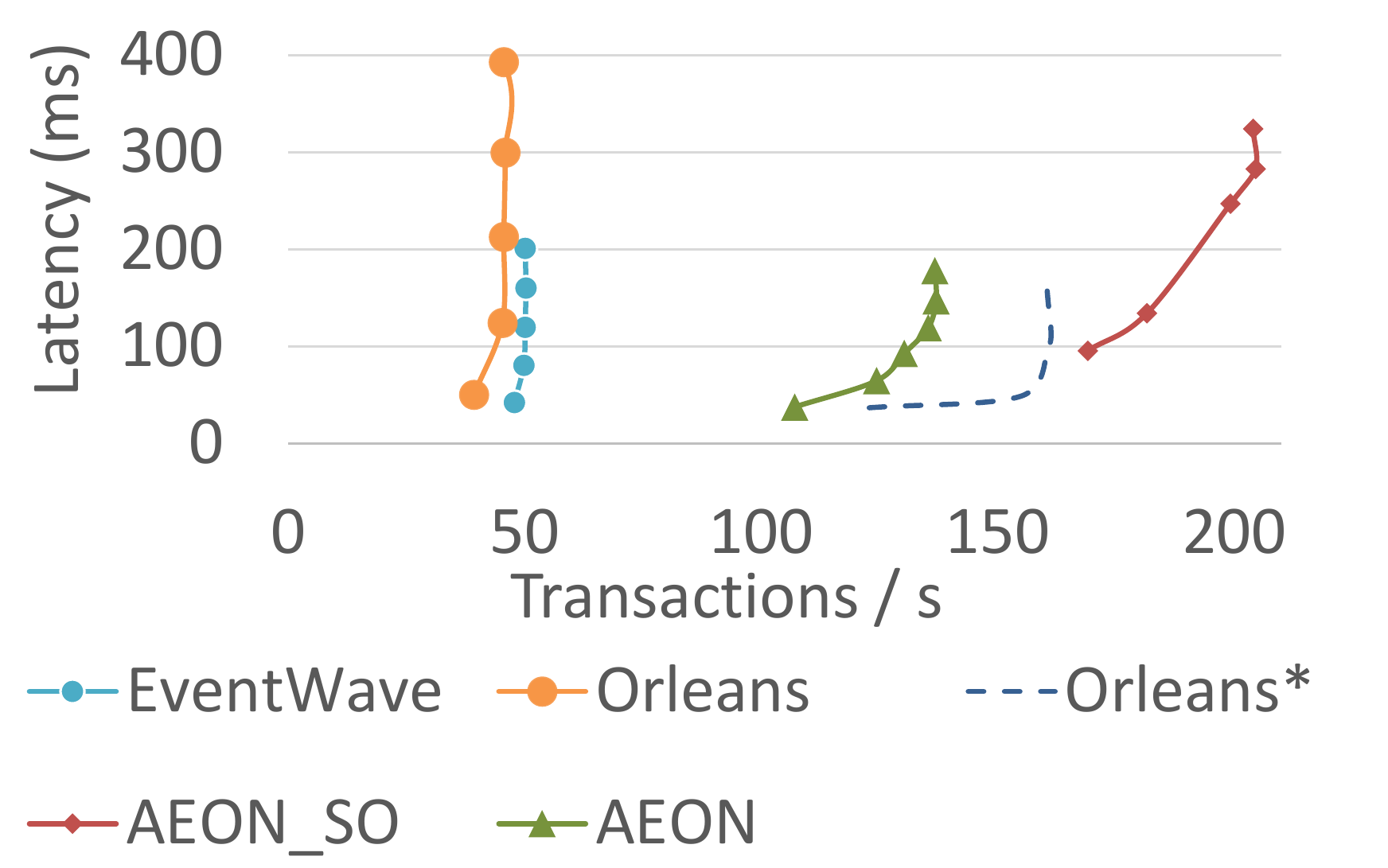}}
  \caption{TPC-C scalability and performance}
\end{figure*}

Because of the parallelism provided by multiple ownership, we observe that
\aeon{'s} performance is 50\% more than \aeonso{} when the number of servers
reaches 16.
More precisely, since \aeonso{} does not have multiple ownership, in order to
access \lstinline{Items} belonging to a given \lstinline{Players},
\lstinline{Room} context needs to be locked.
However, multiple ownership allows both \lstinline{Players} and \lstinline{Room}
contexts to access \lstinline{Items} thus leading to parallel
execution of more events within one \lstinline{Room}.

\vspace{1mm}\noindent\textbf{Performance evaluation.}~
Figure~\ref{fig:Game:Game-performance} plots throughput and latency of the game
application when the number of servers is fixed to 8.
Similar to Figure~\ref{fig:Game:Game-scaleout}, \aeon{} outperforms all other
systems.
As we explained above, optimized \aeon{} exploits allows for more
parallelism and reduces the overhead in communication.
\subsubsection{TPC-C benchmark} \label{sec:scalabilityperformance::tpcc}
The TPC-C benchmark is an on-line transaction processing benchmark.
TPC-C is a good candidate for comparing \aeon{} with its rivals since it has
multiple transaction types with different execution structures.
Observe that transactions in TPC-C are similar to events in \aeon{} and
\eventwave{}.
All of our TPC-C implementations are made fault tolerant through checkpointing.
We note that we used TPC-C solely to stress-test \aeon{}, and evaluate its
performance under high contention. In reality, specifically engineered elastic
distributed databases may be a better fit for serving TPC-C style applications.

The TPC-C benchmark implementation in \aeon{} uses the following context
declarations:\\[3pt]
\scalebox{.8}{
\begin{tabular}{l}
  \lstinline$contextclass WareHouse$ \{\lstinline$set<Stock> s; set<District> d;$\}\\
  \lstinline$contextclass Stock$ \{ ... \}\\
  \lstinline$contextclass District$ \{\lstinline$set<Customer> c; set<Order> o;$\}\\
  \lstinline$contextclass Customer$ \{\lstinline$History h; set<Order> os;$\}\\
  \lstinline$contextclass Order$ \{\lstinline$set<NewOrder> n; set<OrderLine> l;$\}\\
  \lstinline$contextclass NewOrder$ \{ ... \}\\
  \lstinline$contextclass OrderLine$ \{ ... \}\\
\end{tabular}}\\[3pt]
\noindent Since the number of items is fixed (i.e., $100K$) in the TPC-C
benchmark, and does not need elasticity, warehouse and items form a single
context.

Observe that an \lstinline{Order} context has two owners:
\lstinline{District} and \lstinline{Customer}.
In our \aeonso{} and \eventwave{} implementations, and since they should follow
a single ownership structure, the \lstinline{District} context does not own the
\lstinline{Order} context. In other words, the \lstinline{Order} context is
solely owned by the \lstinline{Customer} context.

\noindent A typical approach for evaluating the scalability of a system using TPC-C is to
partition TPC-C by warehouse, and put each warehouse on a single
server~\cite{Cowling2012,Thomson2012}.
But, as pointed out by Mu et al.~\cite{Mu2014}, this approach does not stress
the scalability and performance of distributed transactions (i.e., events in our
programming model) because less than \%$15$ of transactions are distributed.
Therefore, we also partition TPC-C by district similar to Rococo~\cite{Mu2014}.

Similar to the game application, we also implemented two variants of TPC-C in
Orleans:
(i) A version that ensures strict serializability, which we shall name
Orleans throughout this section. This version is implemented by exploiting
the fact that stateful Orleans grains are single-threaded, and we orchestrate
grains in a tree-like structure \`a la \eventwave{}.
(ii) We also implemented a non-strict serializability variant called Orleans*,
in which the strict serializability is not guaranteed to be maintained. We note
that this implementation is only used as a best-case scenario for the
performance of Orleans, and it should otherwise be considered erroneous since it
fails to ensure all the invariants of the TPC-C benchmark.

\vspace{1mm}\noindent\textbf{Scale out.}~ Figure~\ref{fig:TPCC-scaleout}
plots scalability of different systems for the TPC-C benchmark.
In this experiment, we placed one \lstinline{District} (along with its
corresponding \lstinline{Customers}, \lstinline{Orders}, etc.) in each server.
While neither \eventwave{} nor Orleans can scale as the number of servers
increases, we observe that \aeon{} scales up to 4 servers and \aeon{}$_{so}$
scales up to 8 servers.
At this point, the \lstinline{Warehouse} context becomes saturated,
thus \aeon{} and \aeon{}$_{so}$ cannot scale beyond 4 and 8 servers.

More specifically, \aeon{} and \aeonso{} are able to outperform \eventwave{} and
Orleans due to (i) its use of the ownership network to order events, and (ii)
 $\asyncm$ method calls inside events.
As an event (i.e., a TPC-C transaction) finishes its execution in a parent
context, it can continue its execution in a child context by using $\asyncm$
method calls to the child context.
For instance, once a payment transaction finishes its execution in a
\lstinline{Warehouse} context, it calls a method in a \lstinline{District}
context asynchronously, and releases the \lstinline{Warehouse} context.
This allows another event to enter the \lstinline{Warehouse} for execution.

Figure~\ref{fig:TPCC-scaleout} also shows that both Orleans* and \aeonso{}
perform better than \aeon{} when the number of servers reaches 16.
This is because in TPC-C, multiple ownership does not help to
increase the parallelism.
Each \lstinline{District} context owns several hundreds of \lstinline{Customer}
contexts and each \lstinline{Customer} context owns several \lstinline{Order}
contexts.
With multiple ownership, all these \lstinline{Order} contexts are shared by both
 \lstinline{District} context and \lstinline{Customer} contexts.
Consequently, method calls from \lstinline{Customer} contexts to
\lstinline{Order} contexts have to be synchronized by the \lstinline{District}
context, which is the dominator of \lstinline{Customer} contexts.
This leads to the \lstinline{District} context becoming saturated fast.
But, in the single ownership case, the dominators for \lstinline{Customer}
contexts are themselves. Therefore, the \lstinline{District} context does not
become the bottleneck.


\vspace{1mm}\noindent\textbf{Performance evaluation.}~ Figure~\ref{fig:TPCC-performance}
shows TPC-C performance boundaries with 8 servers.
As expected, the throughput of \eventwave{} and Orleans reach maximum with few
clients (i.e., 4-8 clients) and then their latencies skyrocket immediately.
This is due to both implementations failing to handle high contention
at the \lstinline{Warehouse} context properly.
As shown in both Figure~\ref{fig:TPCC-scaleout} and
Figure~\ref{fig:TPCC-performance}, Orleans* outperforms \aeon{} with 8
servers since \aeon{} has to pay extra overhead for strict serializability: 
events will be synchronized by \lstinline{District} context.
\subsection{Elasticity} \label{sec:elasticity:scaling}
As it was explained in \cref{migration}, \aeon{} has several built-in
elasticity policies.
In this section, we solely report our results on evaluating elasticity
capabilities of \aeon{} using the Service Level Agreement (SLA) metric as the
elasticity policy of the game application.

For this experiment, we set the SLA for clients requests to 10ms. 
Therefore, \aeon{} automatically scales out if it takes more than 10ms to handle a client
request.
We placed our clients on 8 m1.large EC2 instances.
Similar to Tuba~\cite{masoud2014a}, we varied the number of clients on each
client machine from 1 to 16 according to the normal distribution. Therefore, at
its peak time, there were 128 active clients in the system.
The game application was deployed on a cluster of m1.small EC2 instances.
To better understand the elasticity capabilities of \aeon{},
we also run the game application with fixed numbers of servers (i.e., 8, 16 and
32 servers).



\begin{figure}[!t]
  \centering
  \begin{tabular}{c}
    \subfloat[Elastic v.s static]{ \label{fig:elastic-game:latency}
      \includegraphics[width=0.95\columnwidth]{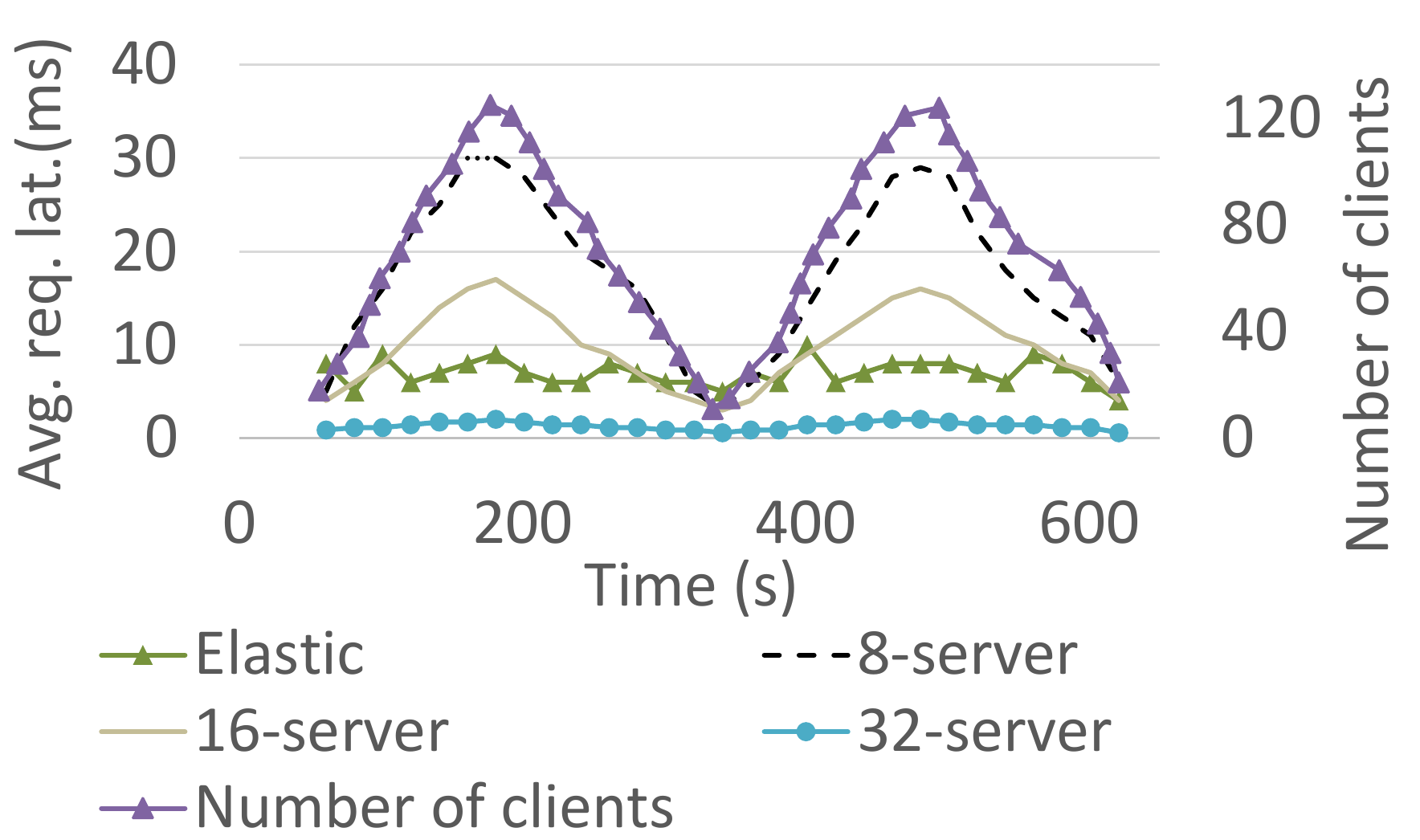}%
    }\\
    \subfloat[Server number]{ \label{fig:elastic-game:sever}
      \includegraphics[width=0.94\columnwidth]{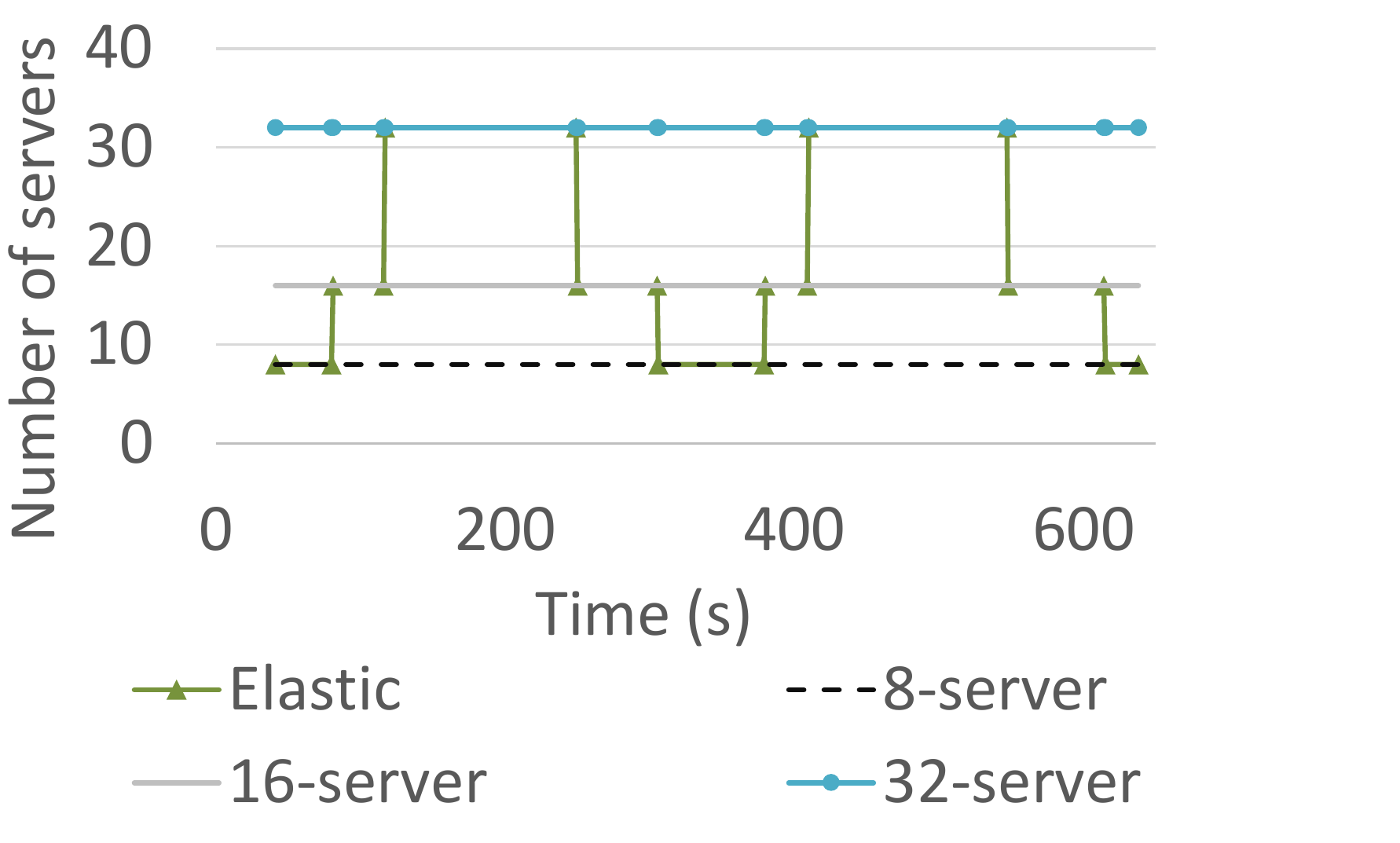}%
    } \\
  \end{tabular}
  \caption{Elastic game application} \label{fig:elastic-game}
  \vspace{5pt}
\end{figure}

Figure~\ref{fig:elastic-game:latency} shows the average request latency that
clients observed, and Figure~\ref{fig:elastic-game:sever} plots the variation of
the number of servers during the experiments.
During peak times, both 8-server and 16-server setups were unable to maintain
the latency below 10ms.
However, elastic and 32-server setups successfully met their SLAs.
Due to migration cost, and fewer servers, clients in the elastic setup observed
a slightly higher request latency.

Table~\ref{table:elasticity} shows the percentage of client requests violating
the SLA and the average number of servers used in each setup.
While both 32-server and elastic setups managed to meet the SLA,
32-server setup did so with 47\% more resources.
Lastly, observe that a (non-elastic) 22-server setup was unable to satisfy
the SLA while the elastic setup fulfilled the SLA with 21.4 servers (on
average).

\begin{table}[!t]
  \centering
  \begin{tabular}{ c |  c  | c }
     Setup & \% of requests > 10ms & Avg. servers \\
    \hline
    8-server & 72.6\% & 8 \\
    \hline
    16-server & 44.2\% & 16 \\
    \hline
    22-server & 20.0\% & 22 \\
    \hline
    32-server & 0.0\% & 32 \\
    \hline
    Elastic & 0.0\% & 21.4 \\
  \end{tabular}
  \caption{Performance and cost} \label{table:elasticity}
\end{table}

\subsection{Migration} \label{sec:elasticity:eManager}
In this section, we first study the effect of migration on the overall
throughput of the system. We then evaluate the throughput of \cservice{} when
performing migration.

\vspace{1mm}\noindent\textbf{Overall throughput.}
We now show the effect of migration for different cases in our game application.
In a first experiment, which we omit for lack of space, we migrated one context
with different sizes up to 100MB.
Clearly, as a context's size increases, the time it takes to transfer from one
server to another increases, but the overall throughput remains stable.

In the second experiment, we migrated different numbers of contexts.
Our experiments were deployed on EC2 with 20 m1.small instances.
We create 20 \lstinline{Room} contexts, one for each host.
We also fixed the size of each context to 1MB, and migrated contexts in order to
determine the accumulated effect of multiple migrations at the same time
-- expected for high workloads.

\begin{figure}[t]
  \centering
      \includegraphics[width=0.95\columnwidth]{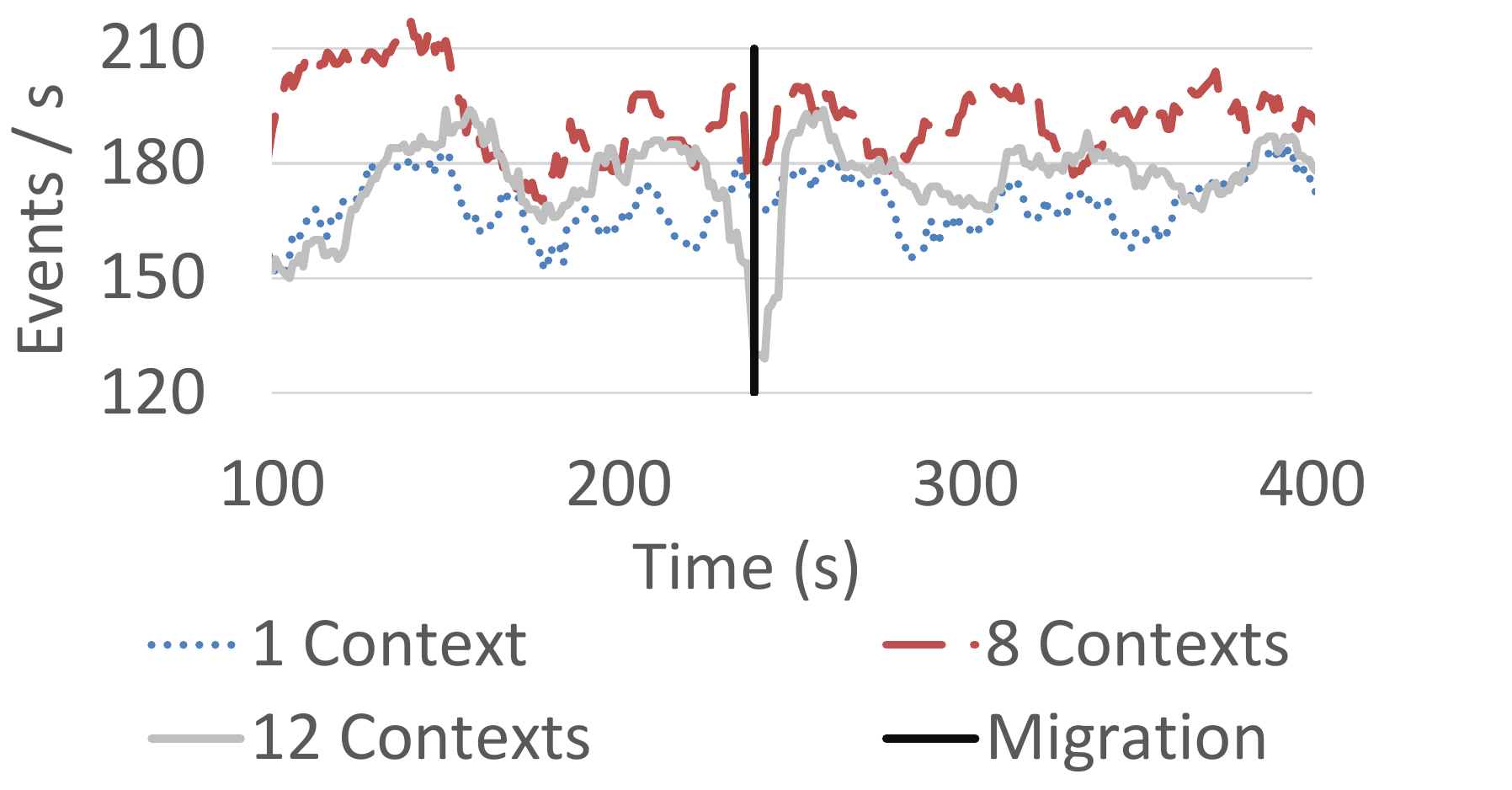}
  \caption{Migrating different number of contexts.}
  \label{fig:piazza_migrate_diff_number}
\end{figure}

Figure~\ref{fig:piazza_migrate_diff_number} shows the overall throughput
variation of the system when migrating different numbers of \lstinline{Room}
contexts.
The mild degradation observed, especially for the case of 12
simultaneous migrations is due to the fact that when a context is
being migrated, requests to it are delayed for the duration
of the migration. In this case more than 50 percent of the contexts
are being moved, which should obviously impact the performance of the
system for a short period of time as shown in the figure.

\begin{figure}[!t]
  \centering
    \includegraphics[width=0.95\columnwidth]{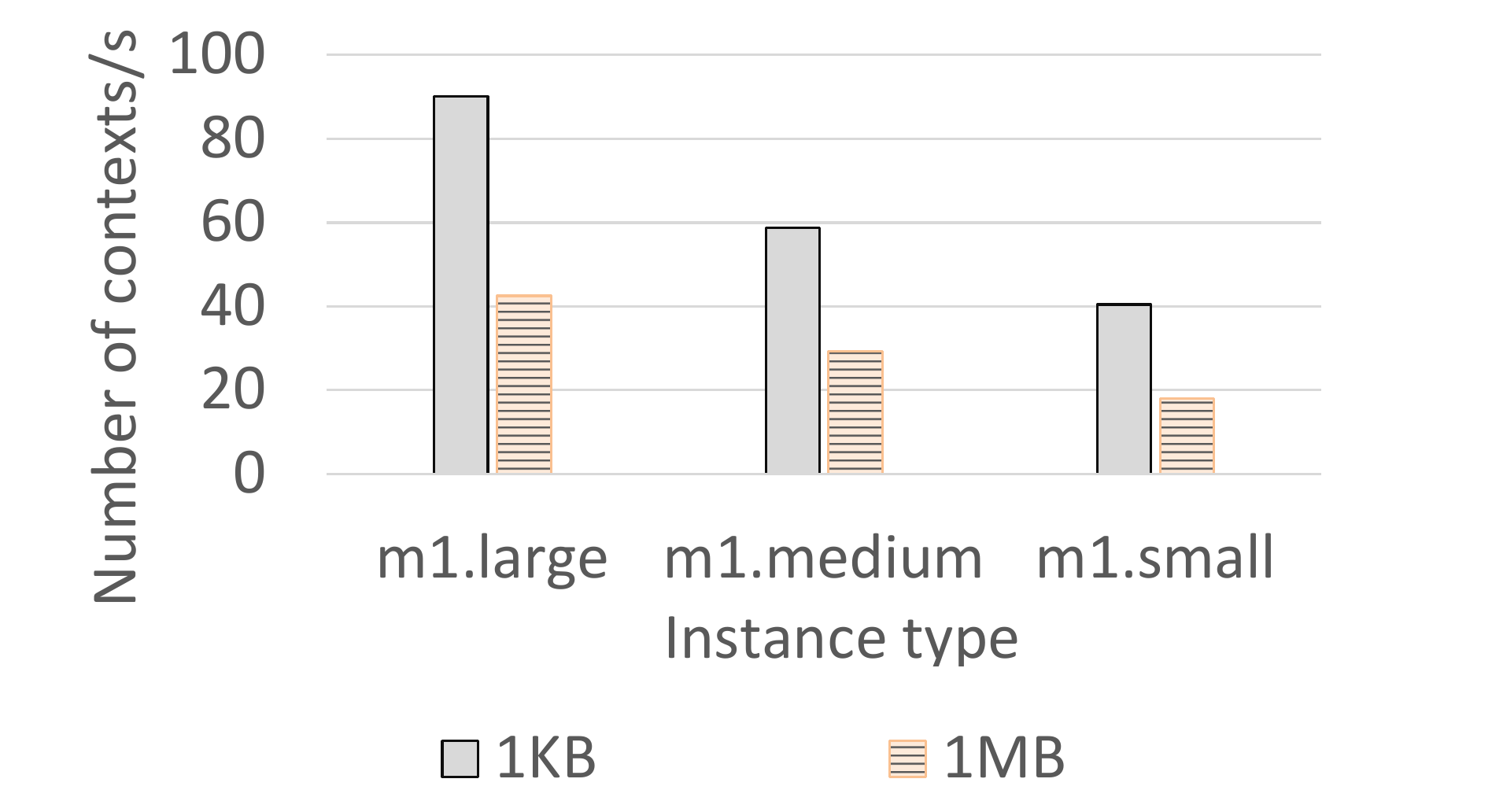}
  \\[-10pt]
  \caption{Max migration throughput on eManager}
  \label{fig:eManager-throughput}
  \vspace{-5pt}
\end{figure}

\vspace{1mm}\noindent\textbf{\cservice{} throughput.}
Finally, we evaluated the maximum throughput of the migration algorithm
introduced in \cref{migration} using a micro benchmark. 
To this end, the \cservice{} moves contexts from one AWS instance to another. 
Figure~\ref{fig:eManager-throughput} shows the \cservice{} throughput with
different context sizes. 
With m1.large instances, the \cservice{} is able to move around 90 small contexts
(i.e., 1KB in size) or 40 large contexts (i.e., 1MB in size) every second.
These numbers are dropped to 60/25 with m1.medium instances, and 40/20 with
m1.small instances. 

We expect that the number of contexts to be much less than the number of objects
for an application. In other words, one context plays the role of a container
for several objects as long as these objects do not require an independent
elasticity policy.
For example, consider the game application. Within a room, there can be several
objects like lights and chairs. These objects can all be included in the
\lstinline{Room} context.
However, in case light object has some non-trivial CPU or memory usage, it
should be treated as a separate context.

\section{Related Work} \label{sec:related}
\vspace{1mm}\noindent\textbf{Distributed programming models.}
The actor model \cite{NoakesWD93,BykovGKLPT11,export:210931} is a
popular paradigm that can be used to develop concurrent applications.
Actors encapsulate state and execute code that can be distributed
across multiple servers.  Actors communicate with each other via
message passing, and there is at most one thread executing in an actor
at all times.  This eliminates the complexities involved in
guaranteeing data race and deadlock freedom.  In that sense, actors
are similar to contexts in our model.  However, it is important to
note that atomicity in actor systems is only given with respect to single 
actors, whilst an event in \aeon{} can atomically modify several
contexts.

Orleans~\cite{BykovGKLPT11} and EventWave~\cite{ChuangSYGKK13},
described in~\cref{sec:evaluation} provide concepts similar to
\aeon{}'s contexts and events. The originality of \aeon{} however resides in the
\emph{ownership network}, which allows us to guarantee
strict serializability, unlike any of these two works, and deadlock freedom
unlike Orleans, while still allowing sharing of state, and
providing opportunities for automatic parallelization and
scale adaptation. EventWave also induces single ownership and limits scalability by invariably synchronizing at a single root node. 

Distributed transactional memory (DTM)~\cite{cloudtm} is a programming paradigm
based on Transactional memory (TM)~\cite{ST95} that allows the programmer to
build strictly serializable distributed applications with sequential semantics
in mind, just as in \aeon.
However, to the best of our knowledge, there is no efficient DTM implementation
that ensures strict serializability and provides implicit elasticity.


Transactors~\cite{transactors} have been proposed as a means to build distributed applications that 
provide strict serializability for events spanning multiple actors.
However, transactors also do not provide support for building applications whose individual actors are distributed across the cloud.
Moreover, there is no support for migrating actors without affecting consistency which is an important
contribution from this work. We remark though that \aeon may be thought of as an extension of transactors to the distributed cloud with support for automatic elasticity.

MapReduce~\cite{mapreduce1} is a functional programming paradigm for the cloud that allows parallelizing computation via two sequential phases: \emph{map}
and \emph{reduce}, to build applications involving huge data sets. However, writing a generic stateful application whose operations
are non-commutative requires extensive synchronization among threads of computation, which is nontrivial to get right in the MapReduce paradigm~\cite{mapreduce2}, 
let alone supporting automatic elasticity.

\aeon{} also shares similarities with models tailored to multi-core
execution environments like Bamboo~\cite{ZhouD10}. Bamboo provides a
data-oriented approach to concurrency, where the programmer implements
tasks, and the runtime system exploits dynamic and static information
to parallelize data-independent tasks.  Bamboo uses locks to implement
a transactional mechanism for data-dependent tasks. Unlike \aeon{},
Bamboo optimizes concurrency for multiple cores; distribution,
migration, and scale adaptation are not considered.

In SCOOP~\cite{NienaltowskiAM03}, objects are considered individual
units of computation and data.  \emph{Separate} calls -- marked by the
programmer -- can be executed asynchronously from the main thread of
execution.  This is similar to the \lstinline{async} calls of \aeon{}.
Similarly, separate calls can only be issued on arguments of a method,
which is SCOOP 's way of controlling what \aeon{} achieves through
multiple ownership and events.  SCOOP is not concerned with distribution
or scale adaptation addressed by \aeon{}.

\vspace{1mm}\noindent\textbf{Distributed programming languages.}
Emerald~\cite{JulLHB87} is an OO distributed
programming language, providing 
locality
functionalities to allow programmers to relocate objects across the
available servers. Unlike \aeon{}, Emerald does not guarantee
atomicity, and synchronization is left to the programmer. 
Moreover,
Emerald was not designed for the cloud, where the existing resources
might be unknown or dynamically allocated. Therefore, Emerald provides
no elasticity. Identical arguments apply to programming languages like 
Erlang and Akka that contrast them from \aeon and render them insufficient
for building complex distributed applications with minimal programming effort.

\vspace{1mm}\noindent\textbf{Transactional key-value stores.}
Elastic databases (e.g., ElasTras~\cite{Das2013},
Megastore~\cite{Baker2011}) are similar to \aeon{}: they partition and
distribute data among a set of servers and provide consistency in the face of concurrent accesses. 
Unlike \aeon, these do not provide a self-contained programming environment for writing generic elastic cloud applications.

\vspace{1mm}\noindent\textbf{Pilot job frameworks.}
A pilot job framework offers dynamic
computational resources to a set of
tasks~\cite{BodeHKLJ00,FreyTLFT02,LuckowLJ10,RaicuZDFW07}.
Applications running on such a framework can be split into a set of isolated tasks
organized either as a ``bag of tasks'' \cite{BodeHKLJ00,FreyTLFT02,LuckowLJ10}
or as a DAG workflow~\cite{RaicuZDFW07}.
In the former case, tasks can execute in any order, while in the latter case,
they should execute in a particular order defined by a DAG.
These tasks are similar to the events of \aeon{}, 
but unlike \aeon{} where events can communicate with each other via contexts,
tasks cannot communicate with each other.

\vspace{1mm}\noindent\textbf{Computation offloading.}
Offloading improves application performance by partitioning it among servers
either at compilation or runtime \cite{GuMGMN04,LiWX01,McGacheyHM11,OuYL06,WangF08}.
Clearly, partitioning at compilation fails to provide elasticity.
Dynamic partitioning, on the other hand, either targets single-threaded
applications, or requires an explicit addition of parallelism in contrast to
\aeon{}.
\section{Concluding Remarks}
\label{sec:conclusion}
We have presented the design and implementation of the \aeon{}
language. \aeon{} provides a sequential programming environment for
the cloud based on the standard paradigm of object-orientation. 
We provide a description of the semantics of \aeon{}, and show that
this semantics exploits parallelism while providing strict
serializability as well as deadlock and starvation freedom.
We have experimentally shown that the \aeon{} runtime system scales as
the number of client requests increases, and it is able to scale-out/in to
provide an economic solution for the cloud. 
In future work we wish to lift some of the restrictions
imposed on the usage of context references in classes and
define a fine-grained elasticity policy language to allow the
programmer control over the locality of contexts and usage of
resources. 


%
\section*{Acknowledgments}
This work was supported by NSF grants \# 1117065, \# 1421910, and \# 1618923, European Research Council grant \# FP7-617805 ``LiVeSoft -- Lightweight Verification of Software'' and German Research Foundation under grant \# SFB-1053 ``MAKI -- Multi-mechanism Adaptation for Future Internet

\bibliographystyle{abbrv}

\bibliography{bib/predef,bib/bsang,bib/msaeida,bib/references,bib/links} 

\end{document}